\documentclass[journal]{IEEEtran}
\usepackage{amsmath}
\usepackage{cuted}
\usepackage{xcolor}
\usepackage{amsmath,amssymb}
\usepackage[compact]{titlesec}   
\usepackage{upgreek}
\usepackage{graphicx}
\usepackage{booktabs}
\usepackage{setspace}
\usepackage{hyperref} 
\hypersetup{
    colorlinks=true,
    linkcolor=blue,
    citecolor=blue,
    urlcolor=blue
}
\usepackage{overpic}
\begin{document}

\title{\huge Modelling Regime Shifts in Continuous Intraday Electricity Markets with State-dependent Hawkes Processes}

\author{Ayoub Jhabli$^{\ast}$$\dag$\thanks{$^\ast$Corresponding author. Email: ayoub.jhabli@\{wur.nl, cern.ch\}}, Tarek AlSkaif$^{\dag}$, Kwabena E. Bennin$^{\dag}$, Bedir Tekinerdogan$^{\dag}$, Axel Naumann$^{\ddagger}$ and Joost M.E. Pennings$^{\S\P}$\\
{\normalsize\itshape $\dag$Information Technology Group, Wageningen University, The Netherlands\\
$\ddagger$European Organization for Nuclear Research (CERN), Switzerland\\
$\S$Marketing and Consumer Behaviour, Wageningen University, The Netherlands\\
$\P$School of Business and Economics, Maastricht University, The Netherlands} 
 
    \thanks{
        This research is fully funded by the Starter Grant awarded to Dr. Tarek Alskaif and Dr. Kwabena Bennin at the Information Technology Group, Wageningen University. It is partly supported by Project HighLO a collaboration between Wageningen University \& Research (WUR) and the European Organization for Nuclear Research (CERN), which aims to identify malicious trading behavior in electricity markets.}}    
\maketitle
\begin{abstract}
The growing importance of intraday trading in Europe, driven by the increasing penetration of renewable energy sources, has led to higher volatility and periods of market stress. Understanding how order flow behaves under varying liquidity conditions requires models that adapt to the state of the market. This paper proposes a multivariate state-dependent Hawkes process to model the flow of orders across different liquidity regimes in the intraday continuous electricity market. The observed market states are identified using a proposed Liquidity Stress Index (LSI), which quantifies stress using the bid–ask spread in the limit order book, order book volume, and mid-price volatility. Transitions between states induce changes in the intensity parameters, resulting in piecewise dynamics that reflect regime-specific order-flow behavior. The model is calibrated on January–March 2024 Dutch intraday continuous cross-border intraday (XBID) hourly order data from EPEX SPOT. The estimated dynamics show that order flow is strongly self-exciting and near-critical, that executed trades trigger new same-side orders while the reverse effect is negligible, and that liquidity stress mainly strengthens transaction self-excitation rather than reshaping the cross-side structure. Robustness checks over the LSI composition and the state partition show that the results are driven by the regime construction itself: alternative index compositions fit within a narrow band, three states fit better than two, and removing the states clearly degrades the fit relative to a single-regime Hawkes benchmark. The results highlight both the value of conditioning order-flow models on liquidity stress and the limits of a single-exponential specification when event frequencies are highly imbalanced.
\end{abstract}
\section{Introduction}
Intraday Continuous (IDC) electricity trading has grown rapidly across Europe, with rising traded volumes and increasing market activity. This development is closely linked to the growing penetration of renewable energy sources, which introduces additional uncertainty into power systems and contributes to higher volatility in intraday electricity prices \cite{eldridge2023rethinking, BIRKELAND20243830}. Because renewable generation, especially wind and solar, is inherently intermittent and difficult to forecast perfectly, market participants increasingly rely on the intraday market to continuously adjust their positions closer to delivery. According to EPEX SPOT, one of the largest European electricity exchanges, 190 TWh was traded in the IDC market in 2024; the Netherlands ranked fourth by volume, after France, Great Britain, and Germany \cite{epex_brochure}. This rapid growth highlights the need to better understand the behavior of the intraday market at a fine temporal scale.

The impact of renewable generation on intraday electricity prices has been documented in several studies. Kulakov and Ziel show that forecast errors in wind and solar production have a nonlinear effect on intraday prices, with solar forecast errors often exerting a stronger influence because of their correlation with peak demand \cite{Kulakov_2021}. Similarly, Goodarzi et al. find that higher renewable forecast errors increase imbalance volumes, which translate into higher spot prices \cite{repec:eee:enepol:v:134:y:2019:i:c:s0301421519304057}. In a broader European context, Cevik and Zhao show that electricity price volatility is highly interconnected across countries, with around 73\% of forecast error variance in national markets driven by cross-border volatility spillovers rather than domestic shocks \cite{Cevik2025Shocked}. Their results further indicate that a higher share of intermittent renewables is a key driver of these international spillovers. Together, these findings suggest that the increased penetration of renewables has made intraday electricity markets more active, more volatile, and more dependent on rapid rebalancing.

This evolution makes it particularly important to study IDC markets at a fine temporal scale, not only through transactions but also through the \emph{order-level} dynamics that shape the Limit Order Book (LOB). Order submissions, cancellations, and executions determine the available liquidity and reflect the strategic interaction of market participants at high frequency. A recent systematic literature review also highlights the growing attention to European intraday markets and emphasizes the need for more research on high-frequency, order-level modeling \cite{birkeland2024research}.

In high-frequency finance, the stochastic arrival of orders and transactions is commonly modeled using Hawkes processes. Introduced by \cite{496a8ed0-5e76-38ab-8d67-eaa3d8cd00d7}, Hawkes processes are self- and mutually-exciting point processes in which past events increase the probability of future events occurring. They have become a standard tool in financial econometrics for modeling irregularly spaced transactions, order arrivals, and systemic risk \cite{bacry2015hawkesprocessesfinance}. In the context of LOBs, Hawkes processes are particularly appealing because they capture the clustering of events and the endogenous feedback mechanisms that characterize modern electronic markets \cite{bacry2015hawkesprocessesfinance, 32b810b1-386f-3c2a-871a-9223972c94dc}.

Applications of Hawkes processes to electricity markets have gained traction in recent years. The work in \cite{jrfm14040161} analyzed market order arrivals in the German IDC electricity market and found strong but short-lived self-excitation together with diurnal variation in the baseline intensity. In \cite{Bergault2024SimulatingAA}, the authors highlight the specific challenges of sparse IDC order books, where irregular liquidity and frequent gaps require tailored point-process models. Building on these insights, \cite{chatziandreou2025optimalexecutionintradayenergy} integrated Hawkes processes with transient price impact models to derive optimal execution strategies for energy trading companies, explicitly accounting for the Samuelson effect and the increase in volatility as gate closure approaches. Overall, these studies demonstrate that Hawkes-type models are well suited to capturing clustering, liquidity stress, and execution dynamics in IDC electricity markets.

Despite the growing interest in IDC markets, most work focuses on realized transactions rather than orders, ignoring the full dynamics of the LOB. Moreover, Hawkes-based studies on IDC markets adopt a single-regime specification — where model parameters remain constant regardless of market conditions — and do not quantify how order arrival intensities vary jointly across liquidity regimes and/or delivery hours. Moreover, the relationship between liquidity conditions and Hawkes parameters remains underexplored for IDC products. In this paper, we address this gap by estimating a state-dependent Hawkes model for Dutch IDC XBID hourly contracts, where the baseline intensities, excitation coefficients, and decay rates all depend on an observed Liquidity Stress Index (LSI) constructed from bid--ask spread, order book volume, and mid-price volatility. Unlike the model proposed in \cite{Morariu-Patrichi04032022}, we explicitly allow the \emph{baselines} to be state-dependent and use LSI to capture the different regimes in every trading session.

The main contributions of this paper are as follows:
\begin{itemize}
  \item \textit{LSI-based regimes:} We construct an interpretable Liquidity Stress Index from bid--ask spread in the LOB, order book volume and mid-price volatility, and use it to define low-, mid-, and high-stress observable regimes in IDC markets.
  \item \textit{Session-specific models:} We estimate a state-dependent Hawkes model for every delivery hour from January 2024 to March 2024 with state-dependent baselines, excitations, and decay rates.
    \item \textit{Stability by design:} We add a spectral-radius penalty with the same target in every state to keep fits away from supercritical values, and derive analytical gradients for estimation.
  \item \textit{Validation and evidence:} We validate the fitted models using time-rescaled residuals and provide empirical evidence on regime-dependent order flow dynamics.
  \item \textit{Robustness:} We compare the retained specification against eleven alternative index compositions, a single-regime Hawkes benchmark, and three alternative state partitions, and show that the results are driven by the regime construction rather than by a particular component choice or discretisation.
\end{itemize}

The remainder of the paper is organized as follows. Section~\ref{sec:electricity} introduces the European power market and describes the mechanics of the IDC market, together with empirical evidence of limit order clustering. Section~\ref{sec:model-training} presents the mathematical formulation of the liquidity stress index LSI and the proposed state-dependent Hawkes process, including the penalized log-likelihood and the analytical gradients. Section~\ref{sec:experiments} describes the training methodology, parameter initialization, and dataset used in the empirical analysis. Furthermore, Section~\ref{sec:experiments} presents and discusses the empirical results, including goodness-of-fit assessments, parameter estimates, and robustness checks over the index composition and the state partition. Finally, the paper is concluded in Section~\ref{sec:concl}.

\section{Background on Electricity Market}
\label{sec:electricity}

This section provides the foundational context and empirical motivation for the modeling framework presented in this study. It begins with an overview of the European short-term electricity trading. We then detail the microstructural mechanics of the Intraday Continuous (IDC) market, including order types and the price-priority execution logic. Finally, we conduct an empirical analysis of Dutch market data to identify key characteristics—such as event clustering, time-to-gate effects, and overdispersion—that motivate the need for a self-exciting modeling approach.
\subsection{Overall Overview}
\begin{figure}[!ht]
\centering
\includegraphics[width=.49\textwidth]{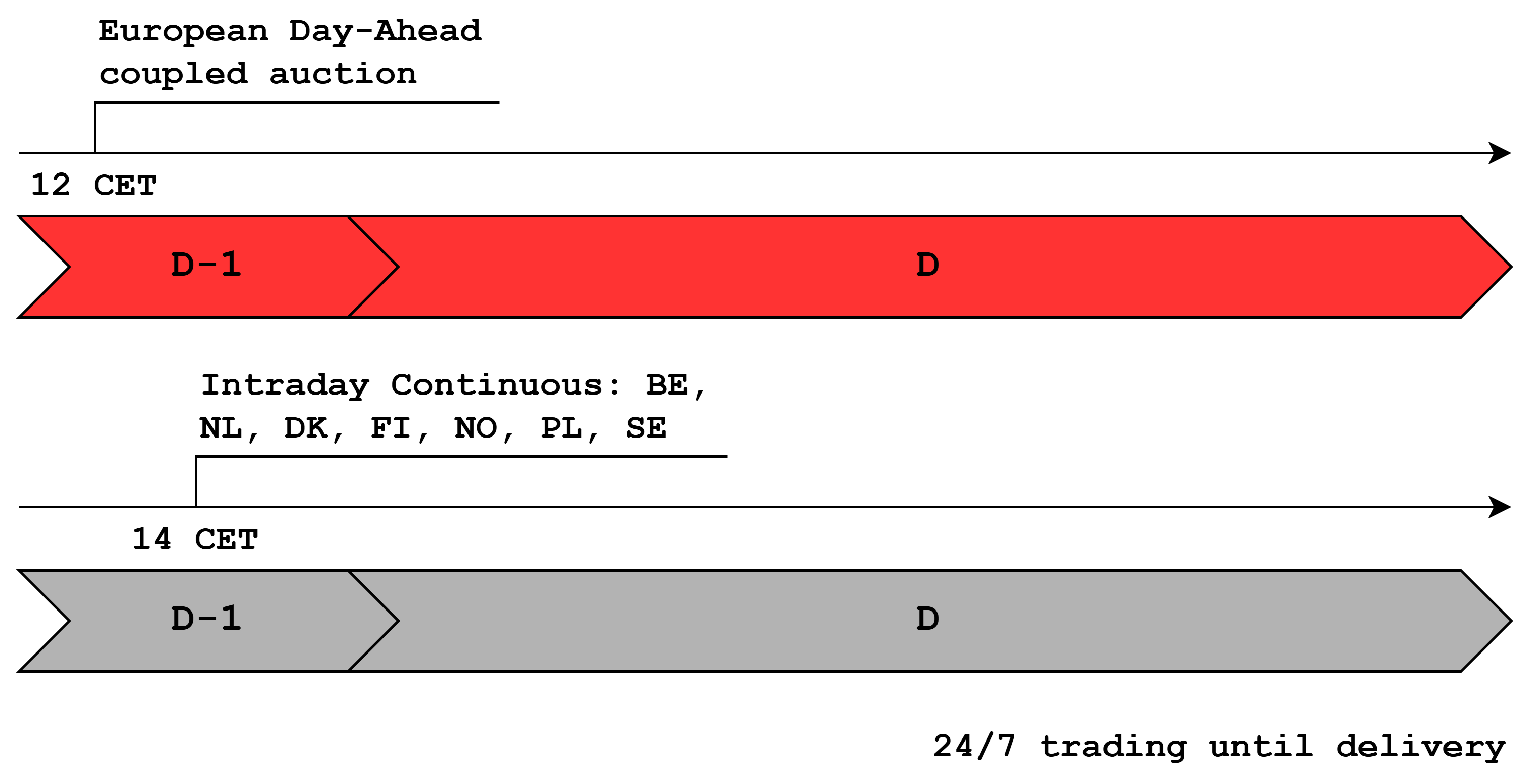}
\caption{EPEX SPOT power trading timeline overview as for February 2025.}
\label{fig:epex_overview}
\end{figure}

Short term electricity trading in Europe takes place in two main markets: the day-ahead and the intraday. Fig.~\ref{fig:epex_overview} shows the trading timeline at EPEX SPOT exchange platform. The day-ahead is auction-based with a uniform price for every 15 minutes and every hour, where electricity can be traded between the different European Union (EU) member states since the introduction of the Single Day-Ahead Coupling (SDAC) in 2015. In most EU countries, the trading starts 45 days in advance and closes one day before delivery at 12:00 CET. Results are published from 12:55 CET onwards. Market participants can balance out their positions in the IDC market that starts at 14:00 CET the day before and continues until delivery starts. In contrast to the day-ahead market, the IDC market does not aggregate all orders into one single price, the transactions in this case happen continuously which generates several prices during the same trading session. Market participants can continuously trade 15-minute, 30-minute, and hourly products at the same time up to 0 minutes before deliveries for Austria, Belgium, France, Finland, and the Netherlands, and 5 minutes for Germany and Switzerland \cite{epex_brochure}. These times refer to the \emph{local} gate closure; cross-zonal trading on the shared order book closes earlier, as detailed in Section~\ref{subsec:idc}.

\subsection{Intraday Continuous Market}
\label{subsec:idc}
EPEX SPOT uses the M7 system from Deutsche Börse for IDC trading. For every message or entry into the LOB, an \emph{Action Code} is attached to classify the event. An added order has \texttt{A (added)}, a matched transaction has \texttt{M (matched)}, and a user-deleted order has \texttt{D (deleted)}. When an order is submitted, M7 first validates whether all entry requirements are met. If the validation is successful, the order is then considered for matching. For local contracts, order actions are performed on the local exchanges. For global contracts, order matching is handled by the Shared Order Book (SOB) module, regardless of the delivery area where the orders were entered. The SOB maintains a consolidated order book for all global contracts.
The order book follows the price–time priority principle: orders with a better price (higher for buy orders, lower for sell orders) are executed first, and among orders at the same price, earlier submissions take precedence. When a new “aggressor” order is submitted with a price that crosses at least one existing order in the book, the system generates one or more transactions. All submitted orders are either deleted or matched before delivery, since any open order remaining at delivery is automatically removed by the system.

Two distinct closures apply to these products. The \emph{cross-zonal gate closure} (GCT) occurs 60 minutes before delivery and terminates the shared order book: from that point onwards an order can no longer be matched against a counterparty in another bidding zone, and the consolidated cross-border book for that contract ceases to exist. Any residual trading in the remaining hour is local-only and is not part of the XBID shared order book. For the XBID hourly products studied in this paper, the GCT is therefore the effective end of the observed trading session.

\begin{figure}[!ht]
\centering
\includegraphics[width=.49\textwidth]{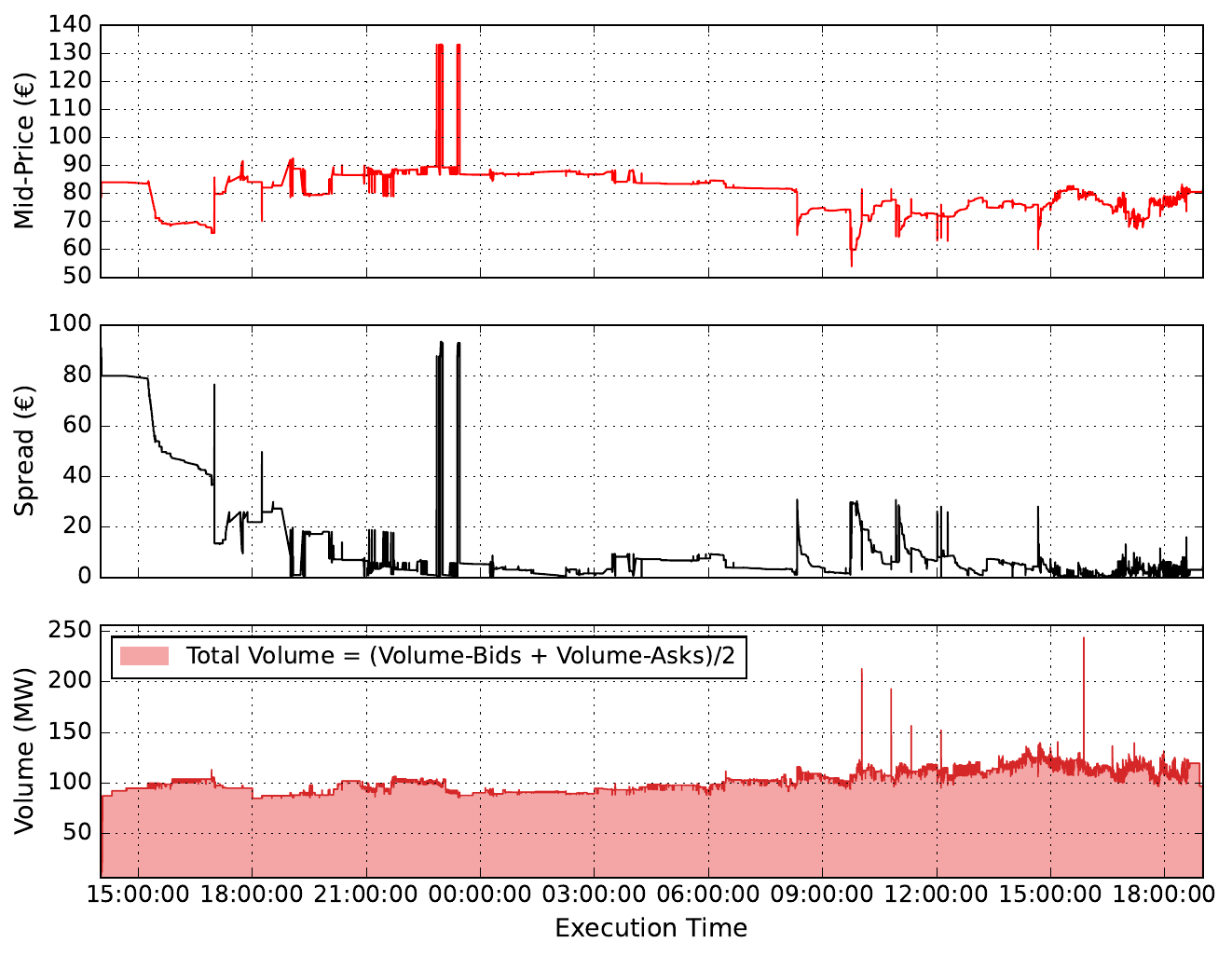}
\caption{Intraday evolution of the XBID market for the delivery hour 20:00--21:00 on the 16th of January 2024. The top panel shows the mid-price trajectory in euros, the middle panel displays the bid-ask spread, and the bottom panel displays the total volume of active orders in the LOB, computed as the average of bid and ask volumes.}
\label{fig:trade_prices_spread_plot_paper}
\end{figure}

Fig.~\ref{fig:trade_prices_spread_plot_paper} illustrates Dutch IDC transaction prices for the XBID hourly contract on the 16th of January for delivery hour 20:00. At the beginning of the trading session, the LOB is sparse and traded volumes are relatively low. As the market approaches gate closure, more accurate forecasts of renewable energy generation, system load, and cross-border capacity become available. Market participants react to this continuous flow of new information, leading to a tightening of the bid–ask spread, increased price volatility, and a noticeable intensification of trading activity. This characteristic intraday behavior is commonly referred to as the Samuelson effect \cite{chatziandreou2025optimalexecutionintradayenergy}.


\subsection{Empirical Analysis}
\label{sec:empirical}

\subsubsection{Clustering, Time-to-gate Effects, and Event-type Asymmetries} 
\begin{figure}[!ht]
\centering
\includegraphics[width=.49\textwidth]{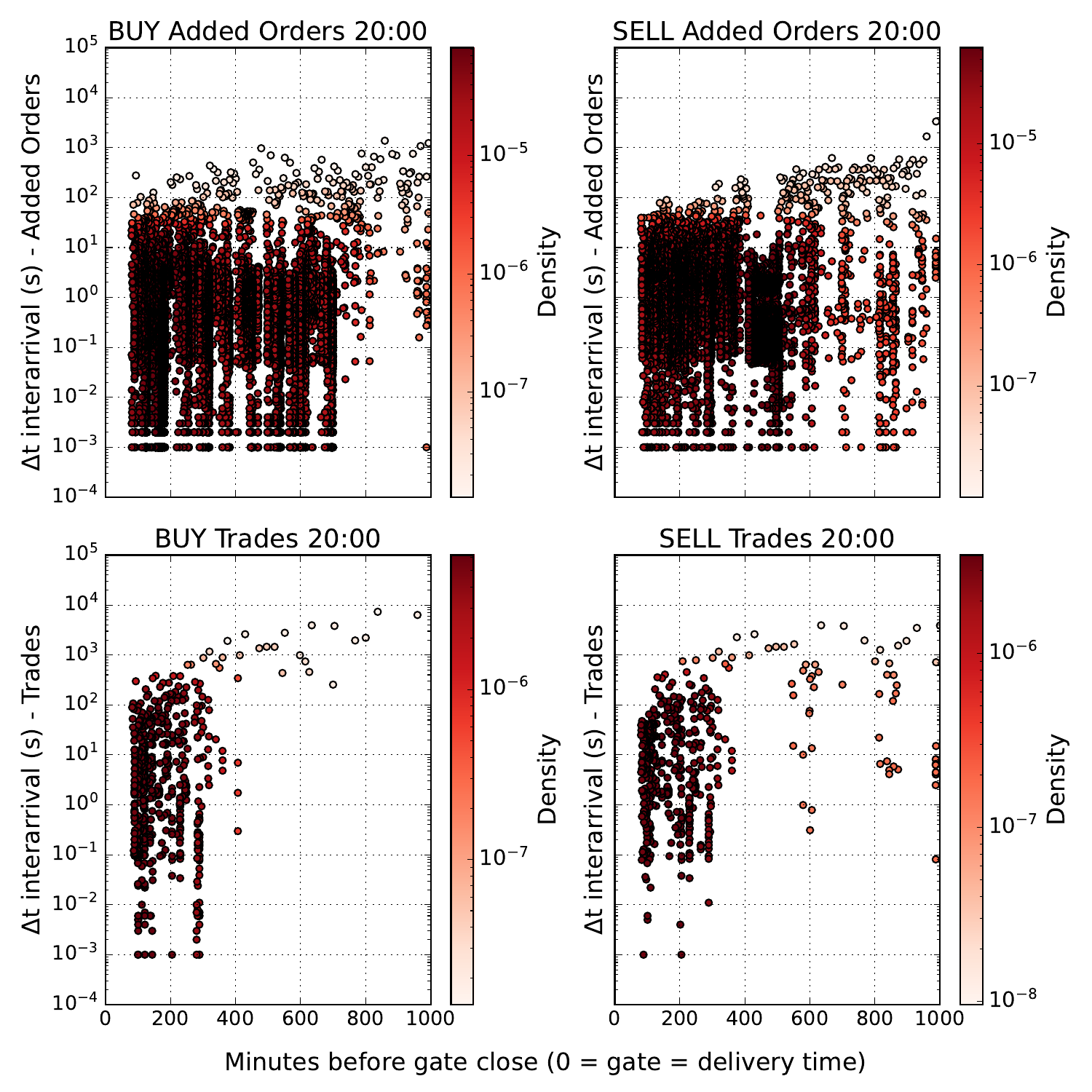}
\caption{Scatter and density plots of order interarrival times against time-to-delivery for an XBID hourly contract (2024-01-16 20:00). The x-axis shows minutes before the delivery time, and the y-axis shows the interarrival time between consecutive orders or trades. Because this is an XBID product, the arrival of orders and trades stops 60 minutes before delivery as explained in \ref{subsec:idc}.}
\label{fig:density_kde}
\end{figure}

Scatter and probability density plots of interarrival times against time-to-gate in Fig.~\ref{fig:density_kde} show that orders arrive in bursts: clusters of very short interarrival times are followed by long inactive periods. These bursts become more pronounced as delivery approaches, indicating strong dependence on time-to-gate. Moreover, the patterns are asymmetric across event types: BUY and SELL added orders, BUY and SELL transactions which can be a full match (\texttt{M}) or partial match (\texttt{P}). 

\subsubsection{Coefficient of Variation and Microstructural Interpretation}

Before specifying a model for order-flow dynamics, we examine the temporal structure of event arrivals through the Coefficient of Variation (CV) of interarrival times:
\begin{equation}
\mathrm{CV}(\Delta t) = \frac{\sigma_{\Delta t}}{\mu_{\Delta t}},
\label{eq:cv}
\end{equation}
where $\mu_{\Delta t}$ and $\sigma_{\Delta t}$ denote the mean and standard deviation of interarrival times, respectively. Under a homogeneous Poisson process, where events arrive independently at a constant rate, the interarrival times are exponentially distributed and $\mathrm{CV}=1$ exactly. Values substantially above one indicate overdispersion, meaning that events cluster in bursts separated by quieter intervals rather than arriving uniformly over time.

\begin{figure}[!ht]
\centering
\includegraphics[width=.49\textwidth]{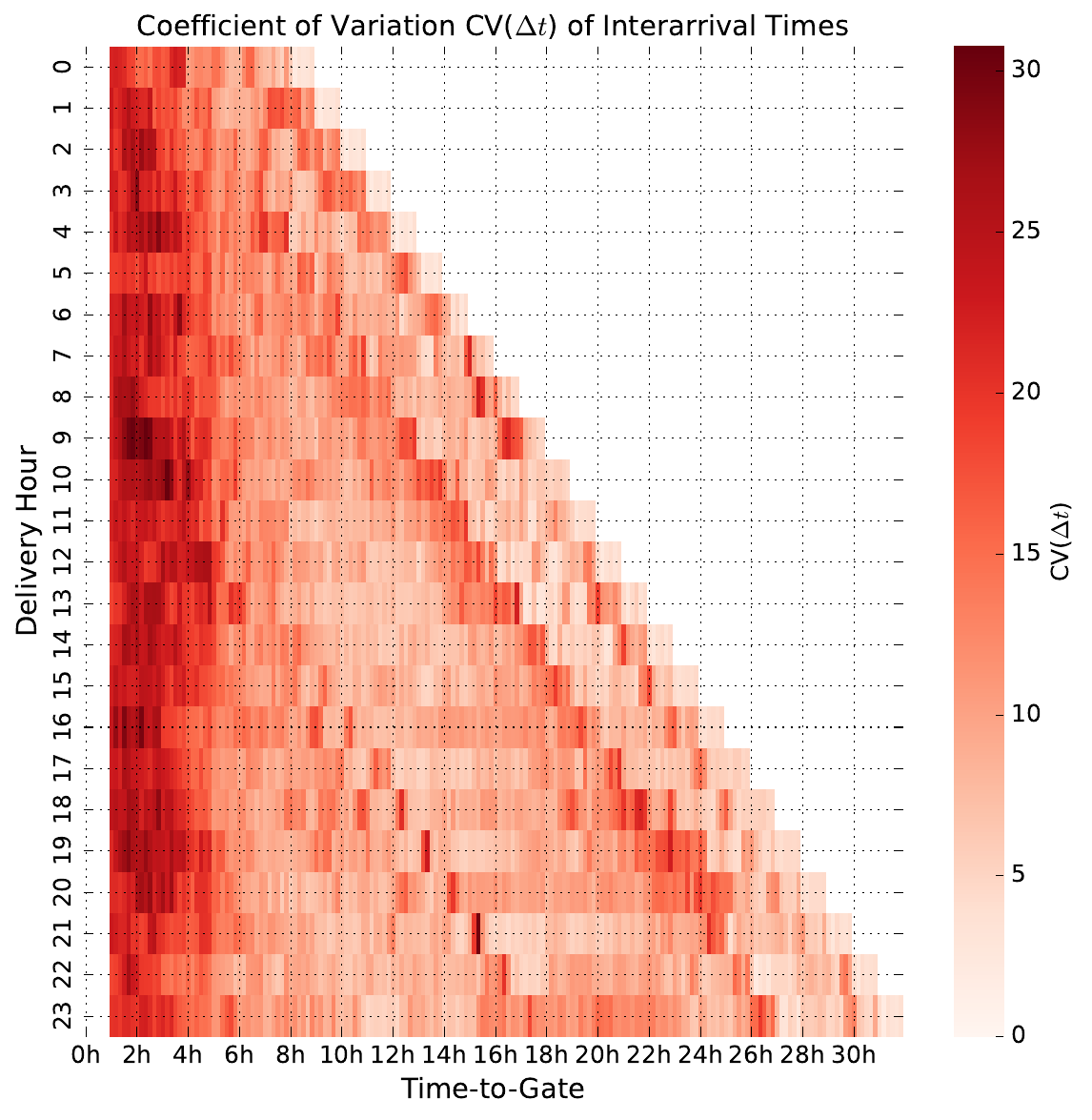}
\caption{Coefficient of variation (CV) of interarrival times across delivery hours and time-to-gate for Dutch XBID hourly contracts in January~2024, computed in 10-minute bins. Values above one indicate overdispersion relative to a Poisson baseline. Burstiness intensifies as the gate closure approaches and varies across delivery hours.}
\label{fig:cv_heatmap}
\end{figure}

Fig.~\ref{fig:cv_heatmap} displays $\mathrm{CV}(\Delta t)$ computed in 10-minute bins across delivery hours and time-to-gate for all Dutch XBID hourly contracts in January 2024. Two patterns stand out. First, CV values are consistently well above one across almost the entire heatmap, confirming that overdispersion is consistent and not confined to specific hours or trading phases. Second, the degree of burstiness is heterogeneous along both axes: it intensifies as the time-to-gate decreases (i.e., as the delivery time approaches).

From a microstructural perspective, such clustering arises naturally when the arrival of one order triggers further submissions from other participants---for instance, algorithmic strategies that react to changes in the order book by rapidly placing new orders, or participants responding to exogenous information such as updated renewable forecasts, congestion signals, or changes in available cross-border capacity. The heterogeneity across delivery hours further suggests that the intensity of these feedback loops varies with the underlying supply-demand conditions of each contract.

These empirical features are fundamentally inconsistent with a Poisson assumption and motivate the use of a \emph{multivariate Hawkes process}, which explicitly models self- and cross-excitation between event types and can accommodate the regime-dependent dynamics suggested by the heterogeneous CV patterns.
\section{Model Description}
\label{sec:model-training}
This section presents the full state-dependent Hawkes process model. We begin by introducing the Liquidity Stress Index (LSI), which serves as the regime variable governing the model dynamics. We then formulate the multivariate Hawkes process with state-dependent parameters. Next, we discuss the branching matrix and the associated stability conditions. We derive the log-likelihood used for estimation, and finally introduce a spectral-radius penalty to encourage stationarity of the models during optimization.

\subsection{Liquidity Stress Index (LSI)}

Liquidity is multidimensional, with no single accepted definition nor an agreed relationship between the proxies used to capture it. Hagemann and Weber \cite{repec:dui:wpaper:1317} identify six dimensions of liquidity in continuous energy markets: tightness, resiliency, depth (or price impact), short-run price volatility, delay and search costs, and trading activity. Three of these are the most consistently used in the literature \cite{kyle1985continuous, mayston2008resiliency}: the bid--ask spread \cite{Lybek2002Measuring, Gabrielsen}, which reflects immediacy costs (tightness); the available volume in the LOB; and the mid-price response to shocks \cite{Xu_2017}, which reflects short-run volatility and resiliency. These proxies show consistent co-movements. Fleming and Remolona \cite{40a095b3-6385-3a05-b033-d615d80e2859} show that around information events, spreads widen, volume surges, and volatility rises together. The role of volume, however, is less settled. \cite{wang_george} find an inverse relationship between volume and the bid--ask spread and a negative relationship between volatility and lagged volume in futures markets, while \cite{Bessembinder_Seguin_1993} show that once volume is split into expected and unexpected components, unexpected volume shocks raise volatility asymmetrically. Trading activity is also close to independent of tightness once active participation is controlled for, suggesting spreads are tied more to trader presence than to raw volume.

Relying on a single proxy to define market regimes may be reductive. One approach is to construct separate models for each proxy, as seen in \cite{Morariu-Patrichi04032022}. The authors utilized the bid--ask spread and queue imbalance to demonstrate how order intensities evolve as a market transitions from wide to tight spreads or from bid-side to ask-side dominance. However, a significant limitation of such independent models is the implicit assumption that order flow dynamics are exclusively driven by a single variable. To capture a more global representation of market liquidity and stress, we propose a composite index that integrates three key proxies: the bid--ask spread, LOB volume, and mid-price volatility. In this framework, the bid--ask spread captures instantaneous liquidity, LOB volume reflects the depth of active interest, and mid-price volatility measures the velocity of short-term price movements. By combining these three metrics, we aim to identify liquidity regimes more comprehensively than models based on isolated variables. Because the components can be combined in more than one way, and because the direction of the volume--stress relationship is disputed, we do not rely on a single construction: we test alternative index compositions, including one in which volume enters with a negative sign, as part of the robustness analysis in Section~\ref{subsec:robustness}.

\subsubsection{LSI construction}
To capture regime shifts in order flow within IDC markets, we define the \emph{Liquidity Stress Index (LSI)} as a composite indicator of market stress based on three microstructural features: the bid--ask spread, the volatility of the mid-price, and the order book volume. Because these features are measured in different units and operate on varying scales, a direct linear summation would disproportionately weigh certain variables and complicate interpretation. To address this, we re-scale the features using $z$-normalization before aggregation. Standardization places all three features on a common, unit-variance scale, and the LSI is simply the average standardized deviation of the three features from their respective means. The LSI is then constructed as:
\begin{equation}
\mathrm{LSI}(t) = \tfrac{1}{3}\Big( z_w(t) + z_{V}(t) + z_{\sigma_{\mathrm{mid-price}}}(t) \Big),
\label{eq:LSI}
\end{equation}
where $z_w(t)$, $z_{V}(t)$, and $z_{\sigma_{\mathrm{mid-price}}}(t)$ denote the standardized spread, total volume in the LOB, and mid-price volatility, respectively. For each feature $x \in \{w,V,\sigma_{\mathrm{mid-price}}\}$, the standardized signal is computed using a fixed historical benchmark:
\begin{equation}
z_x(t) = \frac{x(t)-\mu_x}{\sigma_x},
\label{eq:lsi-zscore}
\end{equation}
where $\mu_x$ and $\sigma_x$ are the mean and standard deviation of feature $x$ estimated over the corresponding contract. In other words, the LSI is used as an ex-post regime-classification device for the fitted contracts, not as a
real-time monitoring signal.


\paragraph{Bid--ask spread.}
Let $P_{\mathrm{bid}}(t)$ and $P_{\mathrm{ask}}(t)$ denote the best bid and best ask prices at time $t$, defined respectively as the highest standing buy price and the lowest standing sell price in the LOB. The bid--ask spread is defined as:
\begin{equation}
w(t)=P_{\mathrm{ask}}(t)-P_{\mathrm{bid}}(t).
\label{eq:spread}
\end{equation}

$w(t)$ measures the instantaneous cost of immediacy and serves as a proxy for market tightness. Larger spread values indicate a less liquid and more stressed market.

\paragraph{Total Volume.}
To capture the displayed liquidity in the LOB, we define the total volume as the average of the cumulative volume available on both the bid and ask sides as:
\begin{equation}
V(t) = \frac{V^{\text{bid}}(t) + V^{\text{ask}}(t)}{2},
\label{eq:totalvolume}
\end{equation}
where
\begin{equation}
V^{\text{bid}}(t) = \sum_{i \in \mathcal{B}(t)} q_i(t),
\label{eq:volume_bid}
\end{equation}
and
\begin{equation}
V^{\text{ask}}(t) = \sum_{i \in \mathcal{A}(t)} q_i(t),
\label{eq:volume_ask}
\end{equation}
with $\mathcal{B}(t)$ and $\mathcal{A}(t)$ denoting the sets of active bid and ask orders in the LOB at time $t$, respectively, and $q_i(t)$ the volume associated with order $i$.

\paragraph{Mid-price volatility.}

To measure recent price volatility, we calculate the standard deviation of the mid-price over the previous 10 minutes. This choice is motivated by the presence of negative prices in European electricity markets, which dismisses the use of logarithmic returns or percentage-based volatility measures. Because log-returns are mathematically undefined for non-positive values, and simple returns become highly unstable near zero, the standard deviation of the raw mid-price provides a more reliable metric for capturing short-term fluctuations in these specific IDC market conditions.

Let the mid-price be defined as:
\begin{equation}
P_{\mathrm{mid}}(t)=\frac{P_{\mathrm{ask}}(t)+P_{\mathrm{bid}}(t)}{2}.
\label{eq:midprice}
\end{equation}
We define the mid-price volatility as:
\begin{equation}
\sigma_{\mathrm{mid-price}}(t)
=
\sqrt{
\frac{1}{N_t}
\sum_{i:\,t_i\in(t-10\text{ min},\,t]}
\left(P_{\mathrm{mid}}(t_i)-\overline{P}_{\mathrm{mid}}(t)\right)^2
},
\label{eq:mid_vol}
\end{equation}
where the mid-price observations are sampled at each message time, $N_t$ denotes the number of such observations in the interval $(t-10\text{ min},\,t]$, and
\[
\overline{P}_{\mathrm{mid}}(t)
=
\frac{1}{N_t}
\sum_{i:\,t_i\in(t-10\text{ min},\,t]}
P_{\mathrm{mid}}(t_i)
\]
is the corresponding local average mid-price. Larger values of $\sigma_{\mathrm{mid}}(t)$ indicate greater short-term price instability and therefore more stressed market conditions.


\subsubsection{Statistical properties of the LSI}

\begin{figure}[!ht]
\centering
\includegraphics[width=.48\textwidth]{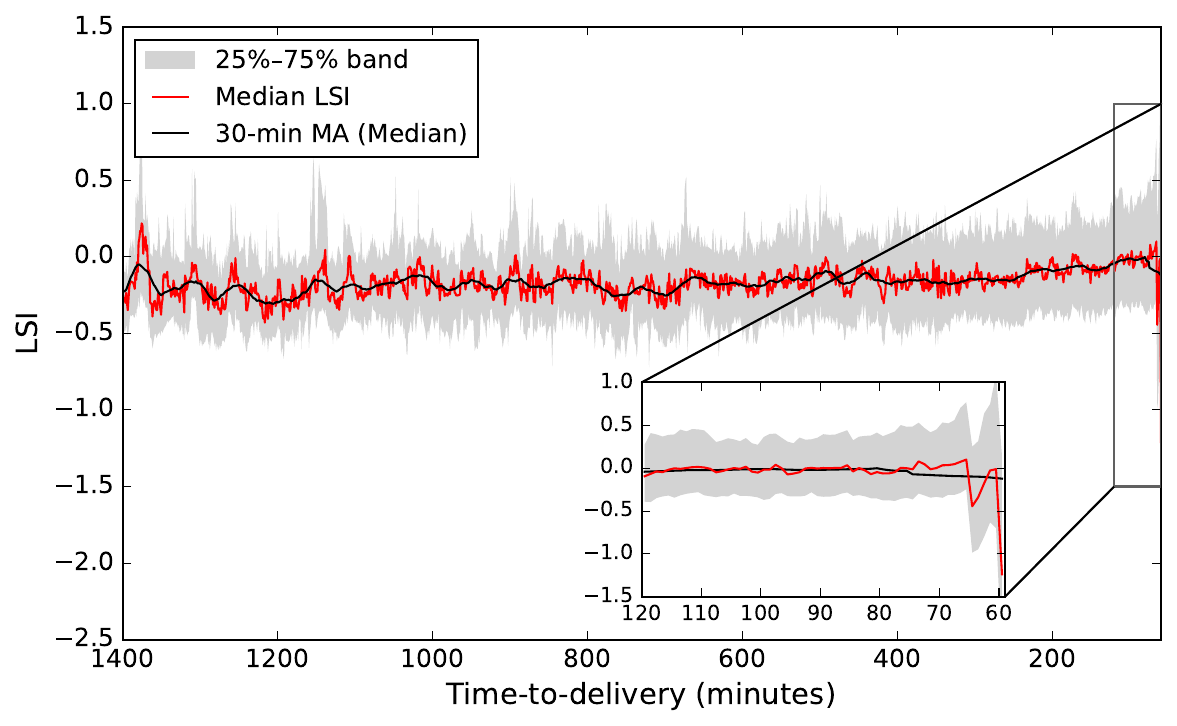}
\caption{Median of the Liquidity Stress Index (LSI) over time-to-delivery per minute bin in January 2024.}
\label{fig:trade_prices_plot_paper}
\end{figure}

Fig.~\ref{fig:trade_prices_plot_paper} shows the median LSI per minute before delivery in January 2024. The index exhibits several peaks in stress and a consistent upward trend throughout the trading session, followed by a sharp decline toward the end. Typically, in each trading session, traded volume increases as time approaches gate closure (i.e., one hour before physical delivery), then drops significantly as the trading session ends and all active orders get closed. 

\subsubsection{State construction and robustness}

The choice of the three components—bid–ask spread, total active volume, and mid-price volatility—and their combination through z-score normalization was partially motivated by their correlation during stressed periods, as discussed earlier. We validate this empirically by defining three states based on the LSI quantiles (0.33 and 0.66), computed per contract so that the thresholds adapt to each contract's own liquidity conditions. These quantiles were selected for two reasons: first, to ensure we have enough samples for each combination of state and event type needed for stable model training; second, to avoid defining high stress as just extreme LSI values, but rather as a distinct market condition.
 
Table~\ref{tab:statewise_summary_compact} presents summary statistics for each state. The three states show clear differences in market conditions: mean bid--ask spreads increase from 9.95 in State~0 to 47.06 in State~2, while mean mid-price volatility rises from 3.21 to 16.20 basis points. Although the standard deviations grow with the state index---reflecting heavier tails in the Stressed regime---the ordering is preserved across different quantiles: the 25th percentile, median, and 75th percentile all increase monotonically from State~0 to State~2, confirming that the state separation is not driven by a few extreme observations but holds throughout the data. This progression confirms that the states capture a range from calm and liquid periods to increasingly stressed market conditions. Total active volume reflects both liquidity and stress levels \cite{40a095b3-6385-3a05-b033-d615d80e2859} and its values are close in the three states. However, it still provides additional and necessary information about the market's microstructure.

\begin{table}[htbp]
\centering
\setlength{\tabcolsep}{3pt}
\caption{State-wise summary statistics of raw features (Dutch XBID, January-March 2024, last four hours of each trading session)}
\label{tab:statewise_summary_compact}
\begin{tabular}{ll ccccc}
\toprule
\textbf{Feature} & \textbf{State} & \textbf{Mean} & \textbf{Std} & \textbf{P25} & \textbf{Median} & \textbf{P75} \\
\midrule
Spread ($w$) & 0 & 9.95  & 28.24 & 2.03  & 5.13   & 12.18 \\
             & 1 & 13.27 & 33.42 & 2.33  & 6.06   & 15.80 \\
             & 2 & 47.06 & 94.97 & 3.66  & 11.57  & 39.90 \\
\addlinespace
Volume ($V$) & 0 & 91.59 & 18.90 & 86.05 & 93.30  & 101.35 \\
             & 1 & 99.38 & 21.00 & 91.65 & 100.20 & 110.10 \\
             & 2 & 103.08& 25.10 & 89.85 & 101.85 & 116.65 \\
\addlinespace
Volat. ($\sigma$) & 0 & 3.21  & 9.50  & 0.54  & 1.43   & 3.05 \\
                  & 1 & 5.20  & 13.41 & 0.85  & 1.95   & 4.33 \\
                  & 2 & 16.20 & 25.83 & 2.16  & 5.76   & 19.63 \\
\addlinespace
LSI          & 0 & -0.51 & 0.21  & -0.60 & -0.46  & -0.36 \\
             & 1 & -0.13 & 0.09  & -0.21 & -0.14  & -0.06 \\
             & 2 & 0.57  & 0.76  & 0.16  & 0.34   & 0.68 \\
\bottomrule
\end{tabular}
\end{table}

Table~\ref{tab:lsi_sweep} reports the sensitivity of the LSI to each component, and, since the thresholds are defined per contract, it also shows the share of contracts in which the swept index crosses each state threshold. Each component is moved from its lower to its upper percentile while the other two are held at their median. Moved from its 5th to its 95th percentile, each component moves the LSI across the lower threshold in almost every contract (97--100\%) and across the upper threshold in most (66\% for spread, 90\% for volume, 99\% for volatility), so none of them is inert. The three span similar ranges overall: 5.1, 5.7 and 7.3 standard deviations from the 1st to the 99th percentile. They differ in where that range sits. For the spread, 68\% of the range lies in the tails, against 39\% for volume and 50\% for volatility. This comes from the skew of the spread, whose standard deviation is inflated by extreme values. For a stress index this is what we want, since the spread is narrow most of the time and widens under stress. Within the interquartile range the upper threshold is reached in a minority of contracts by any single component (28\% for spread, and about half for volume and volatility), so in normal conditions the state is set by the three components jointly rather than by any single one. The LSI and its states provide a basis for modeling the state-dependent Hawkes dynamics. Section~\ref{subsec:robustness} reports a full robustness analysis of both the LSI composition and the state partition.

\begin{table*}[htbp]
\centering
\caption{Sensitivity of the LSI to each component (Dutch XBID, January--March 2024). Each component is moved from its lower to its upper percentile while the other two are held at their median. ``LSI from'' and ``to'' give the resulting index values and ``LSI span'' their difference; ``Component span'' is the component's own range in standard deviations. The last column reports the percentage of contracts in which the swept LSI crosses the lower (33rd-percentile) and upper (66th-percentile) threshold.}
\label{tab:lsi_sweep}
\small
\setlength{\tabcolsep}{5pt}
\begin{tabular}{lccccc}
\toprule
\textbf{Feature} & \textbf{LSI from} & \textbf{LSI to} & \textbf{LSI span} & \textbf{Component span} & \textbf{\% crossing (low / high)} \\
\midrule
\multicolumn{6}{l}{\textit{Sweep from the 1st to the 99th percentile}} \\
Spread ($w$)      & $-0.475$ & $+1.236$ & 1.711 & 5.13$\sigma$ & 100 / 100 \\
Volume ($V$)      & $-1.109$ & $+0.778$ & 1.887 & 5.66$\sigma$ & 100 / 98 \\
Volat. ($\sigma$) & $-0.456$ & $+1.970$ & 2.425 & 7.28$\sigma$ & 100 / 100 \\
\addlinespace
\multicolumn{6}{l}{\textit{Sweep from the 5th to the 95th percentile}} \\
Spread ($w$)      & $-0.337$ & $+0.216$ & 0.553 & 1.66$\sigma$ & 97 / 66 \\
Volume ($V$)      & $-0.687$ & $+0.465$ & 1.151 & 3.45$\sigma$ & 99 / 90 \\
Volat. ($\sigma$) & $-0.363$ & $+0.860$ & 1.223 & 3.67$\sigma$ & 100 / 99 \\
\addlinespace
\multicolumn{6}{l}{\textit{Sweep from the 10th to the 90th percentile}} \\
Spread ($w$)      & $-0.283$ & $+0.048$ & 0.331 & 0.99$\sigma$ & 91 / 43 \\
Volume ($V$)      & $-0.551$ & $+0.323$ & 0.874 & 2.62$\sigma$ & 98 / 80 \\
Volat. ($\sigma$) & $-0.317$ & $+0.526$ & 0.843 & 2.53$\sigma$ & 100 / 92 \\
\addlinespace
\multicolumn{6}{l}{\textit{Sweep from the 25th to the 75th percentile}} \\
Spread ($w$)      & $-0.209$ & $-0.064$ & 0.144 & 0.43$\sigma$ & 78 / 28 \\
Volume ($V$)      & $-0.357$ & $+0.103$ & 0.460 & 1.38$\sigma$ & 94 / 52 \\
Volat. ($\sigma$) & $-0.243$ & $+0.127$ & 0.370 & 1.11$\sigma$ & 94 / 54 \\
\bottomrule
\end{tabular}
\end{table*}

\subsection{Model Formulation}

\subsubsection{Hawkes Process}
The classical one-dimensional Hawkes process as introduced in \cite{496a8ed0-5e76-38ab-8d67-eaa3d8cd00d7} by Alan G. Hawkes, serves as a baseline for our state-dependent formulation. In the classical setting, the system is assumed to evolve under a single, time-homogeneous regime, corresponding to the special case $X=1$ with no state conditioning.

In the one-dimensional case ($E=1$), events are characterized by a single counting process $N(t)$. The conditional intensity of the classical Hawkes process is given by
\begin{align}
\lambda(t)
&=
v
+ \sum_{t_i<t}
  \kappa(t-t_i),
\label{eq:classical_hawkes_1d}
\end{align}
where $v > 0$ denotes the baseline intensity, which can be seen as the component accounting for exogenous events, independent of the past history and $\kappa(\cdot)$ is the excitation kernel that determines how strongly past events influence future events and therefore controlling the strength of the endogenous feedback mechanism. This decomposition illustrates a central advantage of Hawkes processes which is their ability to separate endogenous, history-dependent event generation, from exogenous, history-independent components.

\begin{figure}[!ht]
\centering
\includegraphics[width=.49\textwidth]{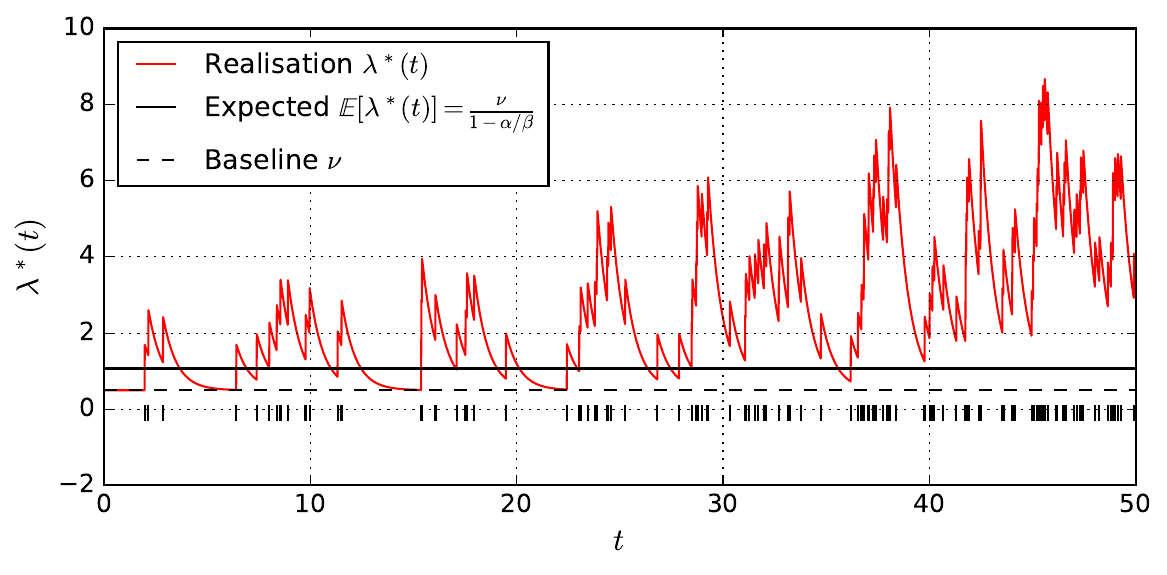}
\caption{Realisation of the conditional intensity $\lambda^*(t)$ of a one-dimensional Hawkes process with exponential kernel, together with its stationary expected value. The process is simulated over the time horizon $T=100$ with baseline intensity $v=0.5$, excitation parameter $\alpha=0.8$, and decay rate $\beta=1.5$. Each jump in $\lambda^*(t)$ corresponds to the arrival of an event and is followed by exponential decay, while the horizontal line indicates the expected intensity $\mathbb{E}[\lambda^*(t)] = v/(1-\alpha/\beta)$.}
\label{fig:intensity_example}
\end{figure}

One of the most widely used kernels in Hawkes processes is the single exponential kernel, defined as:
\[
\kappa(u) = \alpha \, e^{-\beta u},
\]
where $\alpha>0$ controls the magnitude of excitation induced by each event and $\beta>0$ determines the rate at which this excitation decays over time. The exponential form implies a memory that decays rapidly, making it well suited to model situations where the influence of past events diminishes quickly. To illustrate this behavior, Fig.~\ref{fig:intensity_example} presents an example of a simulated intensity trajectory. Each arrival causes an instantaneous upward jump in the intensity $\lambda(t)$, followed by an exponential relaxation back toward the baseline level. 

\subsubsection{State-Dependent Hawkes Process}

As observed in our empirical analysis, the microstructure of the market changes during a single trading session. Therefore, we use a different variant called State-Dependent Hawkes process following \cite{Morariu-Patrichi04032022}.

Let $E$ be the number of event types and $X$ the number of observed LSI-based states.
The observed state process $S(t)\in\{0,1,2\}$ (i.e., low/mid/high stress) is piecewise-constant and aligned to avoid look-ahead (i.e., $S(t^-)$ at event times). For event type $e\in\{1,\dots,E\}$, the conditional intensity is:
\begin{align}
\lambda_e(t)
&= \sum_{x=1}^{X} \mathbf{1}_{\{S(t^-)=x\}}\Big(
  v_{x,e}
  + \sum_{e'=1}^{E}
    \sum_{\substack{t_i<t\\ e_i=e'}}
    \kappa_{x,\,e'\rightarrow e}(t-t_i)
\Big)
\label{eq:intensity}
\end{align}
where 
\begin{itemize}
  \item \emph{Baselines $v_{x,e}$:} Exogenous arrival rate of type $e$ in state $x$, the intensity level in the absence of recent events; it reflects how active the market is under a given liquidity regime.
  \item \emph{Kernel $\kappa_{x,\,e'\rightarrow e}(u)=\alpha_{x,\,e'\rightarrow e}e^{-\beta_{x,\,e'\rightarrow e}u}$:} The memory profile governs how the impact of a past $e'$ event decays over elapsed time $u$; $\beta$ sets the time scale (i.e., larger $\beta$ $\Rightarrow$ faster decay). The long-run cumulative impact equals to
  \[
  \int_0^\infty \kappa_{x,\,e'\rightarrow e}(u)\,du
  \;=\; \frac{\alpha_{x,\,e'\rightarrow e}}{\beta_{x,\,e'\rightarrow e}}
  \;=\; G^x_{e'e},
  \]
  which forms the entry of the state-$x$ branching matrix and links to stability via its spectral radius.
  \item \emph{Excitation coefficients $\alpha_{x,\,e'\rightarrow e}$:} Magnitude of the instantaneous increase in $\lambda_e(t)$ caused by a past event of type $e'$ when the system is in state $x$; large diagonal terms indicate strong self-excitation, off-diagonals capture cross-excitation between event types.
\end{itemize}
\subsection{Branching Matrix and Stability}
For each state $x$, define the branching matrix as:
\begin{equation}
G^x_{e'e}=\frac{\alpha_{x,\,e'\rightarrow e}}{\beta_{x,\,e'\rightarrow e}},\qquad
\rho_x=\rho(G^x),
\label{eq:branching}
\end{equation}
where $\rho(\cdot)$ denotes the spectral radius. Stability (subcriticality) requires $\rho_x<1$. If $\rho_x\ge 1$, the branching representation is supercritical meaning that the expected total number of descendants per event diverges and each arrival has positive probability of having infinitely many descendants.

\subsection{Log-likelihood}
The standard way to estimate the parameters of any
Hawkes process model is to maximize the log-likelihood function. For a given contract, let $\{(t_k, e_k)\}_{k=1}^{K}$ denote the event times and types on $[0,T]$, with observed state path $S(t)$ aligned at $t_k^{-}$.
The log-likelihood is defined as:
\begin{equation}
\mathcal{L}(\theta)
=
\sum_{k=1}^{K}\log \lambda_{e_k}\!\big(t_k\big)
-
\sum_{e=1}^{E}\int_0^{T} \lambda_{e}(t)\,dt,
\label{eq:LL-consistent}
\end{equation}
where $\lambda_{e}(t)$ follows \eqref{eq:intensity} using $S(t^-)$. Each contract is fitted independently.

\subsection{Spectral-radius Penalty}
Let $G^x_{e'e}=\alpha_{x,e'\to e}/\beta_{x,e'\to e}$ and $\rho_x=\rho(G^x)$.
With a common target $\rho^\star=0.99$ applied to all states,
we discourage supercritical fits via
\begin{equation}
\mathcal{P}(\theta)=\sum_{x=1}^{X}\omega\,
\Big(\max\{0,\ \rho_x-\rho^\star\}\Big)^2,\qquad \omega=10^{3},
\label{eq:penalty-consistent}
\end{equation}
and minimize the penalized negative log-likelihood
\begin{equation}
\mathcal{J}(\theta)=-\mathcal{L}(\theta)+\mathcal{P}(\theta).
\label{eq:objective}
\end{equation}
Since the same target applies to every state, no regime ordering is imposed by construction.
The analytical gradients used for L-BFGS-B optimization are provided in Appendix \ref{app:gradients}, and details on the parameter initialization, box constraints, and multi-start procedure in Appendix \ref{app:initialization}.

\section{Experimental Analysis}
\label{sec:experiments}
\subsection{Data and Experimental Setup}
\label{subsec:data}
We estimate contract-specific state-dependent Hawkes process models following \cite{Morariu-Patrichi04032022}, and extend their framework by allowing the baseline intensities, along with the other parameters, to be state-dependent, where the states are constructed using the proposed LSI. For each hourly delivery contract in the January--March 2024 Dutch IDC XBID dataset, we collect the last four trading hours and fit an independent model $\Theta_i$ to that contract's event sequence. Each contract is thus treated as a single session with its own parameter set, yielding one fitted model per contract. 

\begin{figure}[!t]
\centering
\includegraphics[width=.49\textwidth]{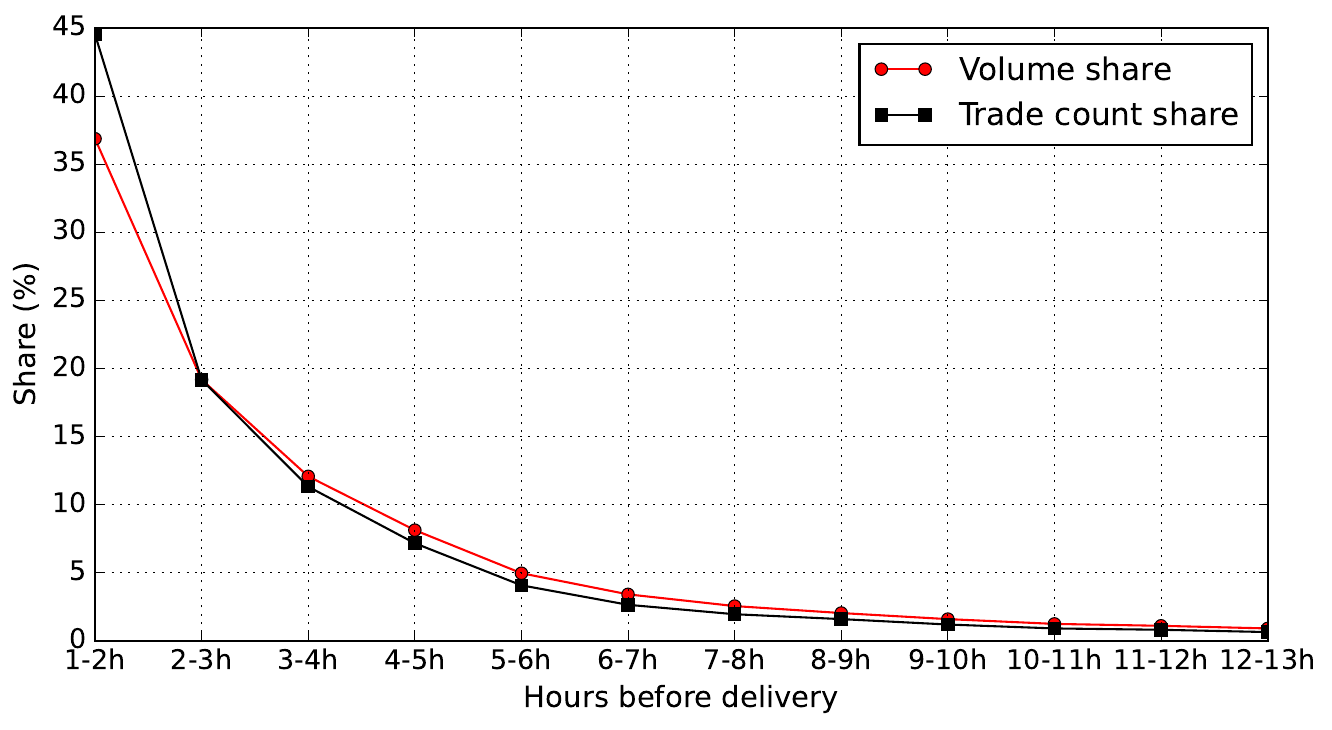}
\caption{Distribution of executed transaction counts and traded volume across time-to-gate intervals for Dutch XBID hourly contracts in 2024. Each bin corresponds to transactions executed between $h$ and $h+1$ hours before delivery. The figure illustrates how both the number of transactions and the traded volume evolve as the market approaches delivery.}
\label{fig:volume_count_share}
\end{figure}

The focus on the final four trading hours is motivated by the concentration of market activity during this period. Fig.~\ref{fig:volume_count_share} shows the distribution of traded volume and transaction counts across time-to-gate intervals, indicating that approximately 69\% of the total traded volume and 76\% of the total number of transactions occur within the last four hours before delivery.

Table~\ref{hourly_events} (Appendix \ref{app:data}) reports the distribution of order events for Dutch IDC XBID hourly contracts from January to March 2024 in the last 4 trading hours of every session.
Columns correspond to the total number of messages and the four event types:
\textbf{$H_{A,S}$} (Added Sell), \textbf{$H_{A,B}$} (Added Buy), \textbf{$H_{T,S}$} (Transaction Sell), and \textbf{$H_{T,B}$} (Transaction Buy). As a reminder, an added order has the action code \texttt{A}, while a \emph{transaction} corresponds to any order getting matched through a full match (\texttt{M}) or a partial match (\texttt{P}). A total of approximately 24.6~million events were used to train our models, where order submissions outnumber executed transactions by a factor of over 8:1.

\subsection{Residual Analysis}
\label{subsec:residuals}

Gooodness-of-fit is assessed using time-rescaled residuals. For each event type $e$,
\begin{equation}
r_{n,e}
=\int_{t_{n-1,e}}^{t_{n,e}} \lambda_e(t)\,dt,
\end{equation}
which should be i.i.d.\ $\mathrm{Exp}(1)$ under correct specification.
For regime-wise checks, we compute state-conditional residuals as:
\[
r^{(x)}_{k,e}
=\int_{t^{(x)}_{k-1,e}}^{t^{(x)}_{k,e}}
\mathbf{1}\{S(t)=x\}\,\lambda_e(t)\,dt,
\]
restricting to type-$e$ events occurring in state $x$; these should also be $\mathrm{Exp}(1)$ \cite{Brown, Morariu-Patrichi04032022}.

We evaluate model adequacy along two dimensions: \emph{distributional fit} (whether residuals follow $\mathrm{Exp}(1)$) and \emph{temporal independence} (absence of serial correlation). Since hourly contracts are fitted independently, diagnostics are computed per contract and then aggregated. For distributional fit, we test whether residuals follow $\mathrm{Exp}(1)$ 
using the Kolmogorov--Smirnov (KS) test at the 5\% significance level for each of the 12 event--state combinations, complemented by the average Wasserstein-1 distance between empirical residuals and the theoretical $\mathrm{Exp}(1)$ distribution. To mitigate the sensitivity of the KS test to large sample sizes, we downsample to $n=100$ residuals per event–state couple before testing. For temporal independence, we apply the Ljung--Box test at the 5\% significance level to assess whether residuals are serially uncorrelated up to 10 lags. Table~\ref{tab:gof_combined} summarises these results, with Panels A and C covering distributional fit and Panel B addressing independence.

\begin{table}[htbp]
\centering
\caption{Goodness-of-fit metrics across contracts.}
\label{tab:gof_combined}
\small 
\begin{tabular}{lrrr}
\hline
\textbf{Event} & \textbf{Low} & \textbf{Normal} & \textbf{Stressed} \\
\hline
\multicolumn{4}{l}{\textit{Panel A: KS test pass rate (\%, $p > 0.05$)}} \\
$H_{A,B}$ & 77.9 & 81.4 & 81.3 \\
$H_{A,S}$ & 71.7 & 78.5 & 78.2 \\
$H_{T,B}$ & 3.6  & 3.9  & 1.5  \\
$H_{T,S}$ & 4.7  & 9.2  & 4.6  \\
\hline
\multicolumn{4}{l}{\textit{Panel B: Ljung--Box pass rate (\%, $p > 0.05$)}} \\
$H_{A,B}$ & 23.6 & 22.5 & 13.5 \\
$H_{A,S}$ & 12.5 & 17.9 & 3.3  \\
$H_{T,B}$ & 68.8 & 78.6 & 76.0 \\
$H_{T,S}$ & 67.5 & 68.8 & 66.7 \\
\hline
\multicolumn{4}{l}{\textit{Panel C: Avg. Wasserstein-1 distance to $\mathrm{Exp}(1)$}} \\
$H_{A,B}$ & 0.170 & 0.175 & 0.174 \\
$H_{A,S}$ & 0.184 & 0.185 & 0.256 \\
$H_{T,B}$ & 0.474 & 0.539 & 0.542 \\
$H_{T,S}$ & 0.598 & 0.487 & 0.486 \\
\hline
\end{tabular}
\end{table}


The results reveal a distinction between order and transaction event types. For order arrivals ($H_{A,B}$, $H_{A,S}$), the KS test is not rejected in 72--81\% of contracts, and Wasserstein-1 distances remain modest (0.17--0.26), indicating that the model captures the marginal distribution of inter-arrival times well. However, Ljung--Box pass rates are low (3--24\%), signaling residual serial dependence that the exponential kernel does not fully absorb. This suggests that order arrivals exhibit clustering at multiple timescales (e.g., rapid sequences of limit order updates) that a single-exponential kernel cannot represent them. The effect is most pronounced in the Stressed state, where Ljung--Box pass rates drop to 13.5\% ($H_{A,B}$) and 3.3\% ($H_{A,S}$), consistent with intensified microstructural activity under high-volatility conditions.

For transactions ($H_{T,B}$, $H_{T,S}$), the pattern inverts. Ljung--Box pass rates are adequate (67--79\%), indicating that the temporal dependence structure is well captured by the Hawkes kernel. Yet the KS test rejects in over 90\% of contracts, and Wasserstein-1 distances are roughly three times larger (0.47--0.60) than for orders. This distributional misfit points to a miscalibration of the compensator which transforms event times into residuals: transactions are likely more bursty than the model predicts, producing an excess of very small residuals alongside occasional long quiescent gaps. A plausible explanation is that transactions are mechanically coupled to the order book state (i.e., a transaction can only occur when a matching order exists) not just past transactions, introducing a dependence on order-side dynamics that the branching structure alone does not fully internalise. Finally, the 8:1 ratio between order additions and transactions noted in Section~\ref{subsec:data}, together with the interarrival statistics in Table~\ref{tab:interarrival_stats} (Appendix \ref{app:inter}), shows that transactions have similar median interarrival times to orders but substantially higher means and skewness, reflecting prolonged quiet periods between execution bursts that Hawkes struggles to reproduce.

\begin{figure}[!t]
\centering
\includegraphics[width=0.47\textwidth]{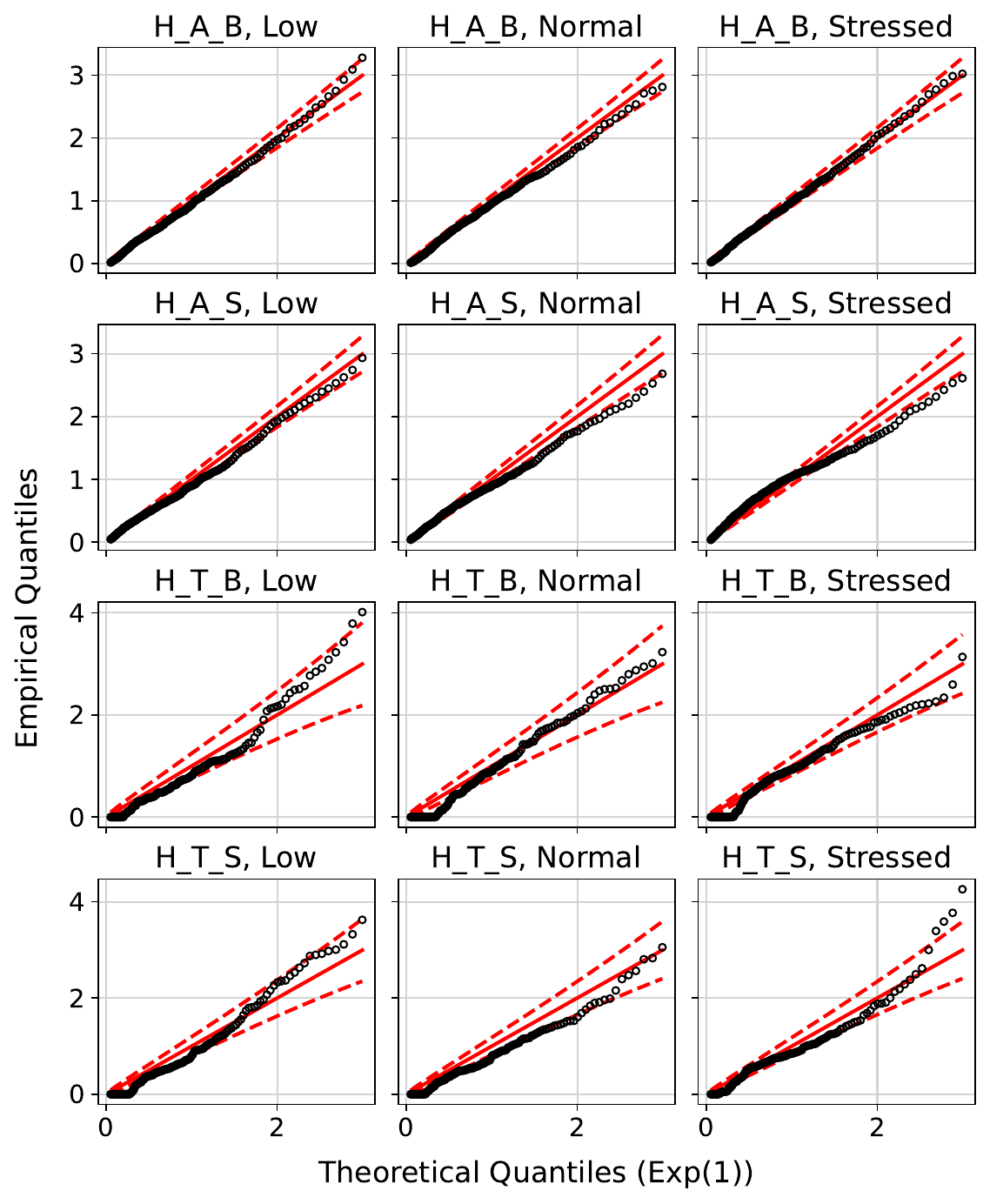}
\caption{Representative QQ-plots of residuals for a single arbitrary chosen hourly contract on the 7th of January 2024 at 5a.m. The solid red line represents the theoretical 45-degree reference, flanked by 95\% asymptotic confidence
bands (dashed red lines).}
\label{fig:qqplots}
\end{figure}

Fig.~\ref{fig:qqplots} displays a representative QQ-plot for a single arbitrary chosen contract. Residuals for order events lie close to the $45^\circ$ line across most quantiles, with upward curvature in the upper tail showing occasional long inter-event periods underrepresented by the model. For the Stressed state, slight downward curvature in the lower tail captures short, high-intensity bursts near delivery. Transaction residuals deviate more substantially but generally stays in the confidence band.

\subsection{State-dependent Baselines and Stability}
\label{sec:baselines}

Table~\ref{tab:baselines} reports the estimated baselines $v_{x,e}$ aggregated over all
2161 fitted contracts. The baselines increase monotonically with the liquidity stress state
for every event type ($v_{\text{low},e} < v_{\text{normal},e} < v_{\text{stressed},e}$), and
the separation is substantial: median within-contract stressed-to-low ratios range from
$1.45\times$ ($H_{T,S}$) to $2.30\times$ ($H_{A,S}$), with paired Wilcoxon signed-rank
$p<10^{-3}$ in all cases. Since each contract acts as its own control, this ordering is not
an artefact of cross-contract heterogeneity. The LSI therefore separates the exogenous
arrival regime clearly, which is the component we make state-dependent as our extension of~\cite{Morariu-Patrichi04032022}.

\begin{table*}[t]\centering
\caption{State-wise baseline intensities $v_{x,e}$ (in units of $10^{-3}$ events/s), aggregated over all fitted contracts ($n=2161$, Jan--Mar 2024). Means with bootstrap 95\% CIs; the last two columns give the paired within-contract Stressed/Low ratio (median) and the Wilcoxon signed-rank $p$-value for $v_{\text{Stressed}}>v_{\text{Low}}$.}
\label{tab:baselines}
\small
\setlength{\tabcolsep}{6pt}
\begin{tabular}{lccccc}
\toprule
Event & Low & Normal & Stressed & Str/Low & $p$ \\
\midrule
$H_{A,B}$ & 20.0 [19.1, 20.9] & 30.2 [29.1, 31.3] & 36.6 [35.3, 38.0] & 1.84$\times$ & $<10^{-3}$ \\
$H_{A,S}$ & 34.4 [33.0, 35.8] & 38.0 [36.7, 39.4] & 80.8 [77.3, 84.3] & 2.30$\times$ & $<10^{-3}$ \\
$H_{T,B}$ & 4.0 [3.8, 4.2] & 4.7 [4.5, 4.8] & 5.8 [5.6, 6.1] & 1.48$\times$ & $<10^{-3}$ \\
$H_{T,S}$ & 4.3 [4.1, 4.5] & 5.0 [4.8, 5.1] & 6.1 [5.9, 6.4] & 1.45$\times$ & $<10^{-3}$ \\
\bottomrule
\end{tabular}
\end{table*}

The picture for the spectral radius is different. Table~\ref{tab:rho} gives $\rho_x=\rho(G^x)$ per state. The fits are near-critical in \emph{every} regime ($\bar\rho \approx 0.89$--$0.92$), and the stress ordering one might expect does not appear: the stressed state is statistically indistinguishable from the low state ($p=0.25$), while the normal state is marginally the highest. Because a single common target $\rho^\star=0.99$ is applied to all states, no ordering is imposed by the penalty, and what we observe is a property of the data. The constraint is slack for the large majority of fits, with the remainder sitting at or just above the target.

\begin{table*}[t]\centering
\caption{Estimated per-state spectral radius $\rho_x=\rho(G^x)$ of the branching matrix ($n=2161$). Means with bootstrap 95\% CIs, share of fits below the stability target, and paired within-contract signed-rank tests. A single common target $\rho^\star=0.99$ is applied to all states, so no regime ordering is imposed: $\rho$ is near-critical in every regime and Stressed is statistically indistinguishable from Low.}
\label{tab:rho}
\begin{tabular}{lccc}
\toprule
State & Mean $\rho$ [95\% CI] & Median & \%\,($\rho<0.99$) \\
\midrule
Low & 0.895 [0.891, 0.899] & 0.916 & 93.8 \\
Normal & 0.922 [0.919, 0.925] & 0.932 & 91.2 \\
Stressed & 0.893 [0.889, 0.896] & 0.910 & 94.5 \\
\midrule
\multicolumn{4}{l}{\footnotesize Paired signed-rank: Normal$>$Low $p<10^{-3}$; Normal$>$Stressed $p<10^{-3}$; Stressed vs Low $p=0.25$ (n.s.).}\\
\bottomrule
\end{tabular}
\end{table*}

Taken together, these two results sharpen the interpretation of the LSI regimes: liquidity
stress reshapes \emph{where} excitation is allocated---raising exogenous arrival rates and
redistributing the off-diagonal structure (Section~\ref{sec:branching})---rather than
raising the overall level of endogeneity, which remains near-critical throughout.

\subsection{Cumulative Excitation Functions}

We visualize excitation dynamics through the \emph{cumulative excitation functions}
for each pair of event types $(e',e)$ and liquidity stress state $x$:
\begin{equation}
\left\lVert \kappa_{x,\,e'\rightarrow e}(\cdot) \right\rVert_{1,t}
:= \int_{0}^{t} \kappa_{x,\,e'\rightarrow e}(s)\,ds.
\label{eq:cumkernel-def}
\end{equation}
For the single-exponential kernels
$\kappa_{x,\,e'\rightarrow e}(t)
= \alpha_{x,\,e'\rightarrow e}\,e^{-\beta_{x,\,e'\rightarrow e}t}$,
the cumulative form admits a closed expression:
\begin{equation}
\left\lVert \kappa_{x,\,e'\rightarrow e}(\cdot) \right\rVert_{1,t}
= \frac{\alpha_{x,\,e'\rightarrow e}}{\beta_{x,\,e'\rightarrow e}}
\bigl(1 - e^{-\beta_{x,\,e'\rightarrow e}t}\bigr),
\qquad t\ge 0.
\label{eq:cumkernel-explicit}
\end{equation}
The plateau value $\lim_{t\to\infty}\lVert\kappa_{x,\,e'\rightarrow e}(\cdot)\rVert_{1,t}
= G^x_{e'e} = \alpha_{x,\,e'\rightarrow e}/\beta_{x,\,e'\rightarrow e}$
corresponds to the entry of the state-$x$ branching matrix defined in~\eqref{eq:branching}.

\begin{figure*}[!t]
\centering
\begin{overpic}[width=.8\textwidth]{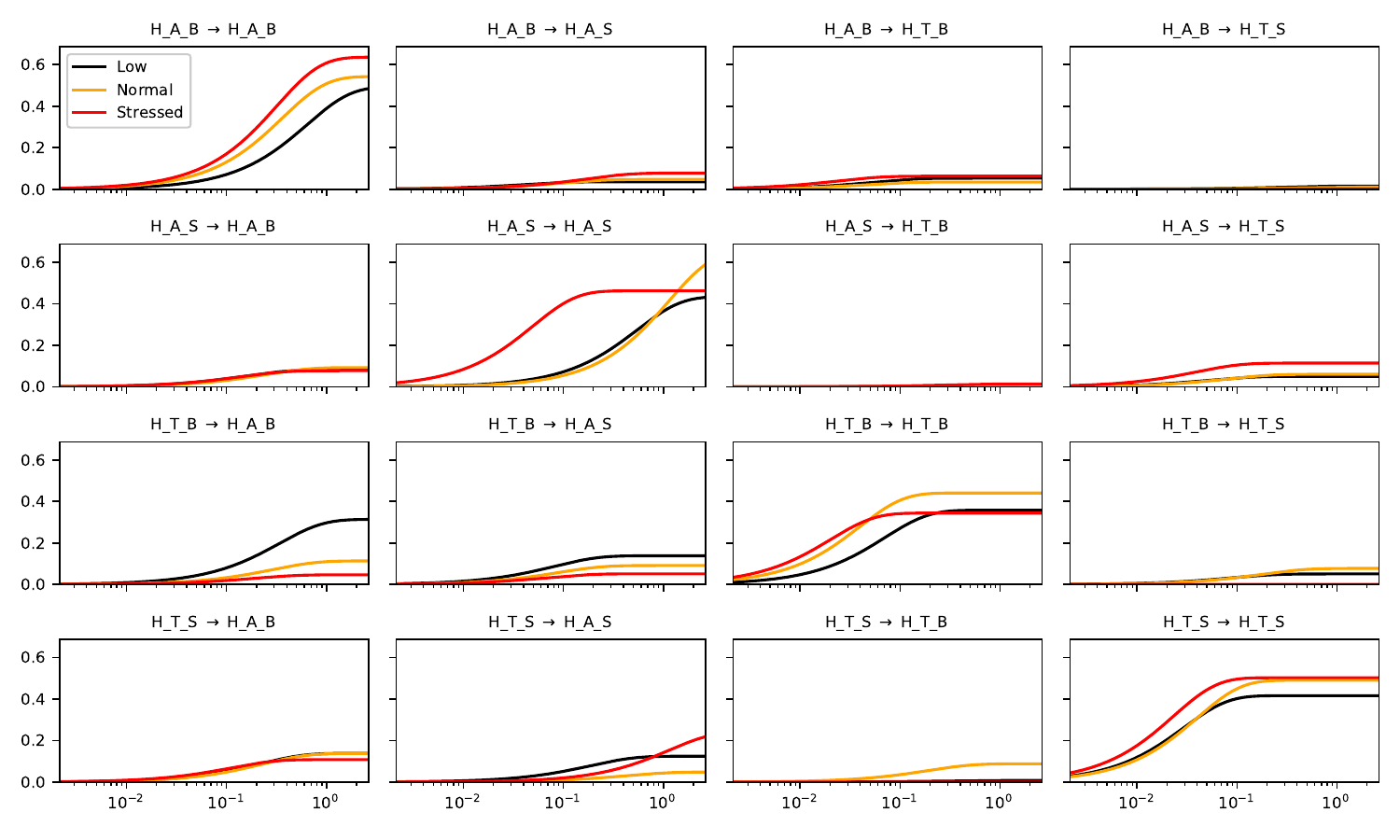}
  \put(50,-.5){\makebox(0,0){\small Time lag $u$ (s)}}
  \put(-1,30){\makebox(0,0){\rotatebox{90}{\small Cumulative excitation $\lVert \kappa_{x,\,e' \rightarrow e}(\cdot) \rVert_{1,t}$}}}
\end{overpic}
\caption{Estimated cumulative excitation functions $\lVert \kappa_{x,\,e' \rightarrow e}(\cdot) \rVert_{1,t}$ for all 16 event-type pairs and LSI states for the delivery contract from 10:00 p.m.\ to 11:00 p.m.\ on January 3, 2024. The time axis is on a logarithmic scale. Qualitatively similar patterns are observed across contracts.}
\label{fig:kernels}
\end{figure*}

\paragraph{Interpretation.}
Fig.~\ref{fig:kernels} displays the cumulative excitation functions for a representative contract. Although each contract is fitted independently, the kernel shapes exhibit consistent qualitative behaviour across the dataset; we therefore discuss the common patterns here and defer the contract-aggregated view to the branching ratio analysis in Section~\ref{sec:branching}.

Several patterns are apparent from the $4\times 4$ panel structure. Along the diagonal, self-excitation is strongest for added orders: $H_{A,B}\!\rightarrow\!H_{A,B}$ and $H_{A,S}\!\rightarrow\!H_{A,S}$ rise quickly and reach clear plateaus, indicating that the submission of a buy (sell) order triggers further buy (sell) order arrivals. Transaction self-excitation ($H_{T,B}\!\rightarrow\!H_{T,B}$, $H_{T,S}\!\rightarrow\!H_{T,S}$) is also present with a different temporal profile: these curves rise more gradually and plateau at longer horizons, suggesting that transaction-to-transaction clustering operates on a slower timescale than order-to-order clustering.

Off-diagonal, cross-excitation is strongest from transactions to orders on the same side of the book. Specifically, $H_{T,B}\!\rightarrow\!H_{A,B}$ and $H_{T,S}\!\rightarrow\!H_{A,S}$ reach clear plateaus, indicating that executed transactions trigger new order submissions. The reverse direction (i.e., orders triggering transactions) is weak, as one would expect: the arrival of a limit order does not mechanically cause an immediate transaction. Cross-side effects (buy$\leftrightarrow$sell) are negligible throughout, indicating limited direct spillover between the two sides of the order book.

The LSI state modulates the magnitude and speed of excitation. In the Stressed regime (red curves), several channels show a steeper initial rise, reflecting faster-decaying kernels (larger $\beta$) that concentrate excitation over shorter horizons. This is especially visible in the transaction-to-order channels ($H_{T,B}\!\rightarrow\!H_{A,B}$, $H_{T,S}\!\rightarrow\!H_{A,S}$), where the Stressed state reaches its plateau earlier than the Low or Normal states. The Low state (black curves) tends to show the most gradual accumulation, consistent with a calmer market environment in which event clustering unfolds more slowly. Across all channels, the curves plateau within approximately $10^{-2}$--$10^{0}$ on the log-scaled time axis, confirming that the memory of the process is short-lived.

\subsection{Branching Ratios}
\label{sec:branching}

The plateau of each cumulative excitation function corresponds to the branching ratio $G^x_{e'e} = \alpha_{x,\,e'\rightarrow e}/\beta_{x,\,e'\rightarrow e}$, which measures the expected number of type-$e$ events triggered by a single type-$e'$ event in state $x$. While the kernel plots above illustrate the temporal dynamics for individual contracts, the branching ratios compress this information into a single scalar per channel and can therefore be meaningfully averaged across contracts to reveal global patterns.

\begin{figure*}[!t]
\centering
\includegraphics[width=.85\textwidth]{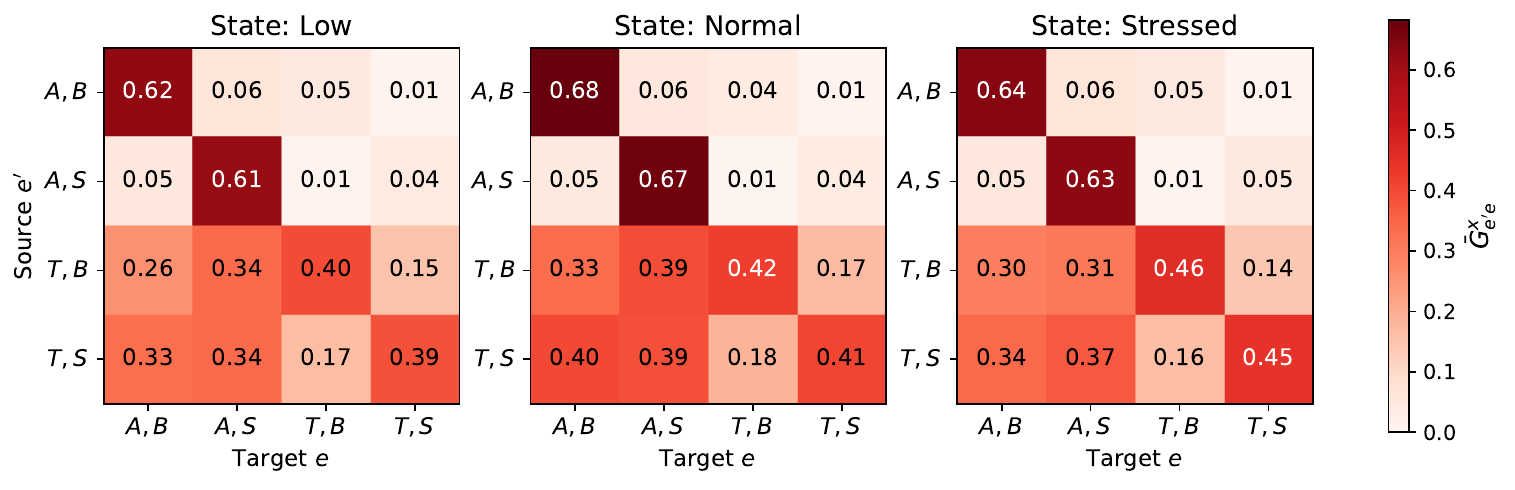}
\caption{Average branching ratio matrices $\bar{G}^x$ for each LSI state, computed across all fitted contracts. Entry $(e', e)$ represents the average expected number of type-$e$ offspring events triggered by a single type-$e'$ parent event.}
\label{fig:branching}
\end{figure*}

Fig.~\ref{fig:branching} presents the average branching ratio matrices $\bar{G}^x$ for each LSI state, computed over all fitted contracts. Three features stand out.

\emph{First}, the matrices are strongly diagonally dominated. Added buy orders excite further added buy orders with an average branching ratio of approximately 0.62--0.68 across all states, and added sell orders similarly self-excite at 0.62--0.67.

\emph{Second}, transaction-to-transaction self-excitation is moderate (0.40--0.46 for $H_{T,B}\!\rightarrow\!H_{T,B}$; 0.39--0.45 for $H_{T,S}\!\rightarrow\!H_{T,S}$), consistent with the slower-rising kernel profiles observed in Fig.~\ref{fig:kernels}. The cross-channel from transactions to same-side orders is also substantial ($H_{T,B}\!\rightarrow\!H_{A,B}$ at 0.26--0.33; $H_{T,S}\!\rightarrow\!H_{A,S}$ at 0.34--0.39), confirming the order replenishment effect identified in the kernel analysis. The reverse direction is negligible---added orders barely excite transactions ($H_{A,B}\!\rightarrow\!H_{T,B}$ and $H_{A,S}\!\rightarrow\!H_{T,S}$ at most 0.05)---so the order--transaction relationship is essentially one-directional: trades trigger new orders, but the arrival of an order does not itself trigger trades.

\emph{Third}, the LSI state modulates the excitation structure mainly along the diagonal rather than off it. Transaction self-excitation strengthens under stress, rising from 0.40 (Low) to 0.46 (Stressed) for $H_{T,B}\!\rightarrow\!H_{T,B}$ and from 0.39 to 0.45 for $H_{T,S}\!\rightarrow\!H_{T,S}$, indicating that trade clustering intensifies in stressed conditions. Cross-side excitation between the two sides of the book, by contrast, is uniformly small and stable: added buy orders excite added sell orders at only 0.06--0.06 across all states, and the reverse channel at ~0.05, so the two sides of the book excite themselves but not each other in every regime. Add self-excitation itself is highest in the Normal state (0.68) and slightly lower in the Low and Stressed states (0.62--0.64), a mild non-monotonicity rather than a directional trend.

\subsection{Robustness Checks}
\label{subsec:robustness}

Finally, we verify that the results above are not driven by the composition of the LSI or by the state discretisation. Holding the data, the event types, and the estimation pipeline fixed, we refit the model under alternative index compositions and alternative state partitions, and compare the fits with the residual diagnostics of Section~\ref{subsec:residuals}: the median per-contract Wasserstein-1 distance ($W_1$) of the time-rescaled residuals to $\mathrm{Exp}(1)$, averaged over the event--state streams (threshold-free; smaller is better), together with the Kolmogorov--Smirnov (KS) and Ljung--Box (LB) pass rates at the 5\% level (larger is better). All refits use one common estimation protocol (a single optimization start with a reduced iteration budget) and identical diagnostics, so the comparison is like-for-like.

\subsubsection{Index composition}
We refit twelve compositions: each standardised component alone ($w$, $V$, $-V$, $\sigma$), the five distinct pairs, the paper index $w{+}V{+}\sigma$ of~\eqref{eq:LSI} and its negative-volume counterpart, and a \emph{no-state} baseline, i.e., a plain multivariate Hawkes process with a single regime and no index. The negative-volume variants are included because thin order books might signal stress as well, so volume can enter the index with either sign. Each composition is refitted per contract on the same events (2161 contracts, January--March 2024). Table~\ref{tab:robust_index} ranks the twelve by $W_1$.

Three observations follow. First, all state-based compositions sit in a narrow band ($W_1$ between 0.293 and 0.355), and every one of them fits better than the no-state baseline ($W_1=0.371$), whose LB pass rate collapses to 0.14 against roughly 0.42 with states: removing the states leaves serial dependence that a single regime cannot absorb. The gap between having a state partition and having none is far larger than any gap among the compositions, so the results are driven by the regime construction itself rather than by a particular component or weighting. Second, no single component drives the results: volume or volatility alone fit about as well as the paper index, spread alone is the weakest state-based composition, and the paper index is close to the best. Third, the ranking is stable over time: ranking the compositions within each month and correlating the rankings across months gives Kendall's $\tau$ between 0.79 and 0.88 (all $p<0.001$). Finally, the order/transaction dichotomy documented in Section~\ref{subsec:residuals} is present under every composition, including the no-state baseline; it therefore originates in the kernel and the sparse transaction process, not in the index.

\begin{table}[htbp]
\centering
\caption{Index-composition ablation ranked by residual goodness of fit (January--March 2024, 2161 contracts). $W_1$ is the median per-contract Wasserstein-1 distance of the time-rescaled residuals to $\mathrm{Exp}(1)$, averaged over event--state streams; KS and LB are mean pass rates at the 5\% level. Compositions are written by their standardised components ($w$: spread, $V$: volume, $\sigma$: mid-price volatility); $-V$ means volume enters with a negative sign. ``No state'' is a single-regime multivariate Hawkes benchmark. Lower $W_1$ and higher pass rates are better; the paper index is highlighted.}
\label{tab:robust_index}
\small
\begin{tabular}{clrrr}
\toprule
Rank & Composition & $W_1$ & KS & LB \\
\midrule
1  & $-V$                       & 0.2934 & 0.347 & 0.420 \\
2  & $-V+\sigma$                & 0.2946 & 0.342 & 0.423 \\
3  & $V$                        & 0.2947 & 0.348 & 0.417 \\
4  & $\sigma$                   & 0.2954 & 0.332 & 0.433 \\
5  & $V+\sigma$                 & 0.2964 & 0.343 & 0.426 \\
6  & $\boldsymbol{w{+}V{+}\sigma}$ \textbf{(paper)} & \textbf{0.3080} & \textbf{0.344} & \textbf{0.435} \\
7  & $w{-}V{+}\sigma$     & 0.3085 & 0.339 & 0.428 \\
8  & $w+V$                      & 0.3124 & 0.346 & 0.429 \\
9  & $w-V$                      & 0.3129 & 0.345 & 0.421 \\
10 & $w+\sigma$                 & 0.3228 & 0.335 & 0.439 \\
11 & $w$                        & 0.3549 & 0.328 & 0.423 \\
12 & no state                   & 0.3707 & 0.314 & 0.141 \\
\bottomrule
\end{tabular}
\end{table}

\subsubsection{State partition}
Holding the index fixed, we then vary only the state partition. We compare four: the paper's per-contract terciles (three states, thresholds at the 33rd and 66th LSI percentiles within each contract); a global tercile partition (the same percentiles computed once over the pooled sample); and two global two-state partitions cutting at the 75th and the 90th percentile. This spans both the number of states and the threshold placement. Table~\ref{tab:robust_states} ranks the four.

Three states fit better than two ($W_1$ of 0.308--0.317 against 0.354--0.387), so the intermediate regime is not redundant: collapsing the market into two states loses fit. Among the four, the paper's per-contract terciles fit best, with the lowest $W_1$ and the highest KS pass rate; because the LSI level varies across contracts, per-contract thresholds adapt to each contract's own liquidity conditions where a pooled cut does not. The ranking is identical in all three months. Under the retained partition, the regimes are also persistent rather than rapidly alternating: the median probability of remaining in the same state at the next event is 0.96 (Low), 0.93 (Normal), and 0.96 (Stressed), with the intermediate state the least persistent.

\begin{table}[htbp]
\centering
\caption{State-partition comparison ranked by residual goodness of fit (January--March 2024). Columns as in Table~\ref{tab:robust_index}. The paper's partition is highlighted.}
\label{tab:robust_states}
\small
\setlength{\tabcolsep}{4pt}
\begin{tabular}{lcrrr}
\toprule
Partition & States & $W_1$ & KS & LB \\
\midrule
\textbf{per-contract terciles (paper)} & \textbf{3} & \textbf{0.3079} & \textbf{0.340} & \textbf{0.436} \\
global terciles                        & 3 & 0.3170 & 0.330 & 0.462 \\
global two-state ($q_{0.75}$)          & 2 & 0.3536 & 0.306 & 0.368 \\
global two-state ($q_{0.90}$)          & 2 & 0.3867 & 0.289 & 0.420 \\
\bottomrule
\end{tabular}
\end{table}

\subsubsection{Stability of the estimated dynamics}
\begin{figure*}[!t]
\centering
\includegraphics[width=.95\textwidth]{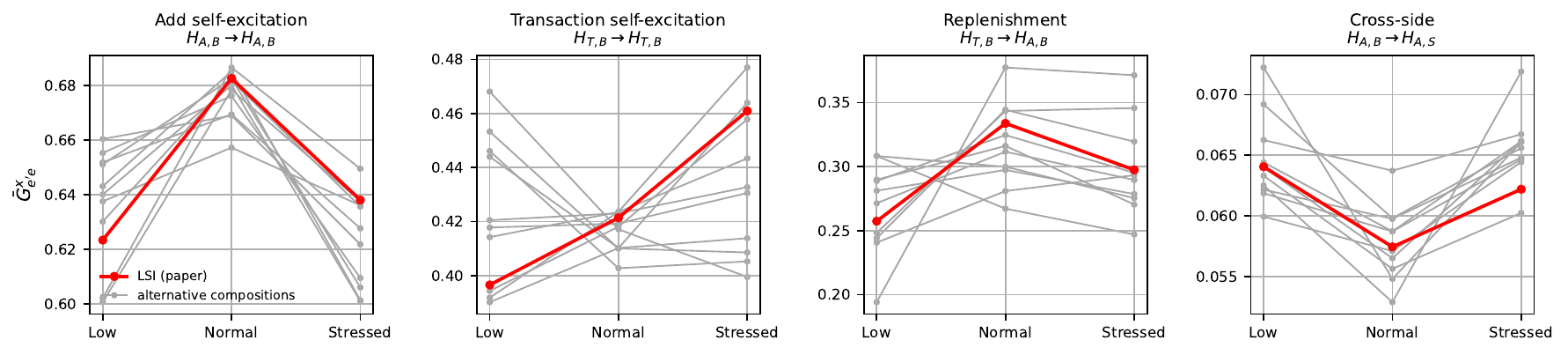}
\caption{Four representative branching channels across LSI states for all eleven state-based index compositions (grey), with this paper proposed LSI in red: add self-excitation ($H_{A,B}\!\rightarrow\!H_{A,B}$), transaction self-excitation ($H_{T,B}\!\rightarrow\!H_{T,B}$), same-side replenishment ($H_{T,B}\!\rightarrow\!H_{A,B}$), and the cross-side channel ($H_{A,B}\!\rightarrow\!H_{A,S}$). The excitation structure is common to all compositions; the cross-side channel is small and flat in every state.}
\label{fig:robust_channels}
\end{figure*}

Fig.~\ref{fig:robust_channels} plots four representative branching channels---add self-excitation, transaction self-excitation, same-side replenishment, and the cross-side channel, analysed in Section~\ref{sec:branching}---across the LSI states for every state-based composition. The channels are tightly grouped: in the Normal state, add self-excitation lies between 0.65 and 0.69 across compositions, transaction self-excitation between 0.39 and 0.42, and same-side replenishment between 0.27 and 0.44, while the cross-side channel stays between 0.05 and 0.07 in every state and every composition. The LSI composition shifts these levels slightly but leaves the excitation structure unchanged.

\subsection{Discussion}

The contract-specific state-dependent Hawkes models reveal both strengths and some limitations with the current framework and the Dutch IDC data. For order arrivals (added buy and sell orders), the model performs well: KS test pass rates remain high, around 70--81\% across all states in January--March, cumulative excitation functions show consistent patterns across contracts, and the average branching ratio matrices $\bar{G}^x$ show interesting interactions between transactions and orders across the different states.


However, the model struggles with transaction event types. One could consider richer kernels (i.e., multi-exponential or power-law) to handle multiple timescales, or a different treatment of transactions. However, the core issue is data sparsity, and richer kernels would worsen the fit rather than improve it. With 4 event types and 3 states, the current single-exponential specification requires 108 parameters per contract (12 baselines plus 96 kernel parameters). Of these, 48 kernel parameters correspond to interactions where a transaction is the source event and are therefore estimated primarily from transaction timestamps. In the Stressed state, the median number of transactions per type per contract is roughly 200, yielding approximately 4 transaction observations per kernel parameter for these pairs --- already at the lower bound for reliable maximum likelihood estimation. While the order event types are abundant (median counts above 1{,}300 per type per state), the transaction-related parameters are bottlenecked by this scarcity. A two-component exponential kernel would double the kernel parameters count to 192, halving this ratio to around 2 observations per parameter and making the trade-related interactions essentially unidentifiable. Pooling events across contracts is not a viable solution either in this market, since IDC hourly products do not roll over: each delivery contract has its own independent trading session with no continuity of the order book from one contract to the next.

The robustness checks in Section~\ref{subsec:robustness} support this diagnosis. The order/transaction dichotomy is essentially unchanged under every index composition and every state partition, so it originates in the kernel and the sparse transaction process rather than in the regime construction. The same checks also confirm that the regime construction itself carries the fit: removing the states leaves pronounced residual autocorrelation, while the differences between alternative index compositions, and between reasonable partitions, are small in comparison.

Despite these limitations, several directions remain promising. First, the state process is currently exogenously defined from the LSI; future extensions could endogenise state transitions and jointly infer their dynamics with event intensities, for instance through a Markov-Modulated Hawkes process~\cite{wu2021markovmodulatedhawkesprocessessporadic}. Second, the discrete three-state structure may oversimplify the continuum of liquidity conditions; continuous or semi-continuous latent-state formulations could capture smoother regime changes. Third, explicit cross-market dynamics between day-ahead and intraday markets or between hourly and quarter-hourly products in IDC could provide additional insights. Finally, an asymmetric model specification (i.e., using richer kernels for order-to-order interactions where data is abundant while regularising or simplifying the transaction-related ones) could offer a pragmatic path to improving the overall fit without worsening the sparsity problem.

\section{Conclusions}
\label{sec:concl}
This study proposes a state-dependent Hawkes process to model order-flow dynamics for hourly XBID products in the Dutch IDC electricity market. Each hourly contract is fitted independently, with observed states defined using a liquidity stress index based on the bid--ask spread in the LOB, order book volume and mid-price volatility. The models capture the marginal arrival dynamics of order events well, revealing near-critical self-excitation, economically interpretable cross-event patterns, and variation across liquidity regimes: exogenous intensities rise with stress, and transaction self-excitation intensifies, while the overall level of endogeneity remains near-critical in every regime. The analysis also identifies a fundamental limitation: the large imbalance between order and transaction frequencies in this market makes it difficult for a standard Hawkes specification to jointly model both event types, as the sparse transaction data cannot support reliable kernel estimation even with a single-exponential specification. Residual diagnostics further indicate that a single-exponential kernel does not fully absorb the serial dependence in order arrivals, particularly under stress. Robustness checks show that these conclusions are not tied to the particular index or discretisation: eleven alternative index compositions fit within a narrow band, a single-regime Hawkes benchmark fits clearly worse, and the proposed per-contract three-state partition fits best among the alternatives considered. These findings highlight both the value of conditioning point-process models on liquidity stress indicators and the need for richer model specifications that can accommodate the heterogeneous frequencies of different event types in electricity intraday markets.

\bibliographystyle{IEEEtran}
\bibliography{references}


\appendices
\setcounter{section}{0}
\renewcommand{\thesection}{\Alph{section}}
\section{Gradients for Single-Exponential Kernels}
\label{app:gradients}
For each $(x,e'\!\to\! e)$, define the (state-filtered) history transforms:
\begin{align}
R_{x,e'\to e}(t)
&=\!\!\sum_{\substack{t_i<t\\ e_i=e'}} 
    e^{-\beta_{x,e'\to e}(t-t_i)},
\label{eq:R}\\
\widetilde{R}_{x,e'\to e}(t)
&=\!\!\sum_{\substack{t_i<t\\ e_i=e'}} 
    (t-t_i)\,e^{-\beta_{x,e'\to e}(t-t_i)}.
\label{eq:Rtilde}
\end{align}
Let the exposure time of state $x$ be:
\begin{equation}
\mathcal{T}_x=\int_0^{T}\mathbf{1}\{S(t)=x\}\,dt.
\label{eq:exposure-consistent}
\end{equation}
Then the gradients of the log-likelihood are:
\begin{equation}
\begin{aligned}
\frac{\partial \mathcal{L}}{\partial v_{x,e}}
&= \sum_{\substack{k:\, e_k=e\\ S(t_k^{-})=x}}
   \frac{1}{\lambda_e(t_k)}
   \;-\; \mathcal{T}_x .
\end{aligned}
\label{eq:dLdv-consistent}
\end{equation}
\begin{equation}
\begin{aligned}
\frac{\partial \mathcal{L}}{\partial \alpha_{x,\,e'\rightarrow e}}
&= \sum_{\substack{k:\, e_k=e\\ S(t_k^{-})=x}}
   \frac{R_{x,e'\rightarrow e}(t_k)}
        {\lambda_e(t_k)} \\
&\quad - \int_{0}^{T}
   \mathbf{1}\{S(t)=x\}\,R_{x,e'\rightarrow e}(t)\,dt .
\end{aligned}
\label{eq:dLdalpha-consistent}
\end{equation}
\begin{equation}
\begin{aligned}
\frac{\partial \mathcal{L}}{\partial \beta_{x,\,e'\rightarrow e}}
&= - \sum_{\substack{k:\, e_k=e\\ S(t_k^{-})=x}}
   \frac{\alpha_{x,\,e'\rightarrow e}\,
         \widetilde{R}_{x,e'\rightarrow e}(t_k)}
        {\lambda_e(t_k)} \\
&\quad + \int_{0}^{T}
   \mathbf{1}\{S(t)=x\}\,
   \alpha_{x,\,e'\rightarrow e}\,
   \widetilde{R}_{x,e'\rightarrow e}(t)\,dt .
\end{aligned}
\label{eq:dLdbeta-consistent}
\end{equation}

\paragraph{Penalty gradients}
Let $u^{(x)}$ and $v^{(x)}$ denote the left and right Perron eigenvectors of $G^x$,
normalized such that $(v^{(x)})^\top u^{(x)} = 1$.
When $\rho_x > \rho^\star$, the gradient of the penalty term
$\mathcal{P}(\theta) = \sum_{x}\omega \big(\max(0, \rho_x - \rho^\star)\big)^2$
with respect to the model parameters is:
\begin{equation}
\begin{aligned}
\frac{\partial \mathcal{P}}{\partial \alpha_{x,\,e'\rightarrow e}}
&= 2\,\omega\,(\rho_x-\rho^\star)\;
   \frac{u^{(x)}_{e'}\,v^{(x)}_{e}}{\beta_{x,\,e'\rightarrow e}},\\[3pt]
\frac{\partial \mathcal{P}}{\partial \beta_{x,\,e'\rightarrow e}}
&= -2\,\omega\,(\rho_x-\rho^\star)\;
   u^{(x)}_{e'}\,v^{(x)}_{e}\,
   \frac{\alpha_{x,\,e'\rightarrow e}}{\beta_{x,\,e'\rightarrow e}^{2}},
\end{aligned}
\label{eq:pen-grads-consistent}
\end{equation}
and $\partial \mathcal{P}/\partial v_{x,e}=0$. 
If $\rho_x \le \rho^\star$, the penalty gradient is zero.
\paragraph*{Proof.} For each state $x$, the branching matrix is $G^x_{e'e} = \alpha_{x,\,e'\rightarrow e}/\beta_{x,\,e'\rightarrow e}$, whose spectral radius $\rho_x$ is the largest real eigenvalue satisfying $G^x v^{(x)} = \rho_x v^{(x)}$ and $(u^{(x)})^\top G^x = \rho_x (u^{(x)})^\top$. Differentiating the eigenvalue equation with respect to $G^x_{ij}$ gives: \[ E^{ij} v^{(x)} + G^x \frac{\partial v^{(x)}}{\partial G^x_{ij}} = \frac{\partial \rho_x}{\partial G^x_{ij}} v^{(x)} + \rho_x \frac{\partial v^{(x)}}{\partial G^x_{ij}}, \] where $E^{ij}$ is the elementary matrix with a single $1$ at $(i,j)$. Premultiplying by $(u^{(x)})^\top$ and using $(u^{(x)})^\top G^x = \rho_x (u^{(x)})^\top$ and $(u^{(x)})^\top v^{(x)} = 1$ yields: \[ \frac{\partial \rho_x}{\partial G^x_{ij}} = u^{(x)}_i\,v^{(x)}_j. \] Applying the chain rule to $G^x_{e'e}=\alpha_{x,\,e'\rightarrow e}/\beta_{x,\,e'\rightarrow e}$ then gives: \[ \frac{\partial \rho_x}{\partial \alpha_{x,\,e'\rightarrow e}} = \frac{u^{(x)}_{e'}\,v^{(x)}_{e}}{\beta_{x,\,e'\rightarrow e}}, \qquad \frac{\partial \rho_x}{\partial \beta_{x,\,e'\rightarrow e}} = -\,u^{(x)}_{e'}\,v^{(x)}_{e}\, \frac{\alpha_{x,\,e'\rightarrow e}}{\beta_{x,\,e'\rightarrow e}^2}. \] Finally, by differentiating $\mathcal{P}_x = \omega (\rho_x - \rho^\star)^2$ for $\rho_x > \rho^\star$, we obtain the expressions in \eqref{eq:pen-grads-consistent}.

\section{Data}
\label{app:data}
To train and validate our models, we utilize LOB data from the Dutch Intraday Continuous (XBID) market spanning from January to March 2024 provided by EPEX SPOT, which is summarized in Table \ref{hourly_events}.
\begin{table}[!ht]
\centering
\caption{Hourly order event counts (Dutch XBID, January-March~2024)}
\label{hourly_events}
\small
\begin{tabular}{rrrrrr}
\hline
\textbf{H} & \textbf{total} & \textbf{$H_{A,B}$} & \textbf{$H_{A,S}$} & \textbf{$H_{T,B}$} & \textbf{$H_{T,S}$} \\
\hline
0  & 707434 & 274155 & 348376 & 41869 & 43034 \\
1  & 837780 & 331260 & 418901 & 43137 & 44482 \\
2  & 935797 & 375929 & 469371 & 43842 & 46655 \\
3  & 950564 & 360477 & 498304 & 45704 & 46079 \\
4  & 824343 & 333181 & 398911 & 46069 & 46182 \\
5  & 959566 & 389721 & 478518 & 45805 & 45522 \\
6  & 1056837 & 481709 & 481427 & 46695 & 47006 \\
7  & 1079880 & 419937 & 556983 & 51655 & 51305 \\
8  & 1249237 & 513330 & 622046 & 57870 & 55991 \\
9  & 1227040 & 533448 & 571207 & 62613 & 59772 \\
10 & 1219924 & 476016 & 608382 & 68727 & 66799 \\
11 & 1235659 & 547454 & 545076 & 76055 & 67074 \\
12 & 1108521 & 490894 & 477152 & 73672 & 66803 \\
13 & 1109684 & 459253 & 514089 & 71589 & 64753 \\
14 & 1172639 & 499873 & 540825 & 69223 & 62718 \\
15 & 1245870 & 522508 & 603245 & 62204 & 57913 \\
16 & 1148778 & 487709 & 534680 & 66577 & 59812 \\
17 & 1097540 & 487035 & 484688 & 63968 & 61849 \\
18 & 1174105 & 557048 & 494688 & 61964 & 60405 \\
19 & 1084916 & 443973 & 535400 & 53250 & 52293 \\
20 & 919221  & 393020 & 425648 & 52095 & 48458 \\
21 & 881859  & 351843 & 430072 & 48424 & 51520 \\
22 & 828158  & 353774 & 378300 & 46341 & 49743 \\
23 & 639495  & 252467 & 311548 & 37588 & 37892 \\
\hline
Total & 24694847 & 10336014 & 11727837 & 1336936 & 1294060 \\
\hline
\end{tabular}
\end{table}

\section{Inter-arrivals}
\label{app:inter}
Table \ref{tab:interarrival_stats} summarizes the distribution of interarrival times for each event type. Median interarrival times are comparable across all four types (approximately 0.16--0.21 seconds), suggesting a similar typical pace of activity. However, mean interarrival times diverge sharply: order additions average around 4 seconds, while transactions average roughly 35 seconds. This discrepancy reflects the clustered nature of executions, which arrive in rapid bursts followed by prolonged quiet periods, whereas order additions flow more continuously throughout the session.

\begin{table}[!ht]
\centering
\caption{Summary statistics of interarrival times (seconds) by event type for Dutch XBID hourly contracts in January 2024.}
\label{tab:interarrival_stats}
\begin{tabular}{lrrrr}
\toprule
{} &
\textbf{$H_{A,B}$} & \textbf{$H_{A,S}$} & \textbf{$H_{T,B}$} & \textbf{$H_{T,S}$}\\
\midrule
Mean     & 3.9697  & 3.6982  & 34.4986  & 35.6974  \\
Std      & 30.6346 & 29.5344 & 183.8534 & 192.7686 \\
25\%     & 0.0580  & 0.0570  & 0.0001   & 0.0001   \\
Median   & 0.2090  & 0.1990  & 0.1600   & 0.2030   \\
75\%     & 1.4860  & 1.5580  & 11.5700  & 12.7379  \\
Skewness & 53.1589 & 53.5758 & 19.2642  & 19.4875  \\
\bottomrule
\end{tabular}
\end{table}

\section{Initialization and Optimization}
\label{app:initialization}
A good initialization is important to help the optimization of \eqref{eq:objective} start from realistic values and reach a stable solution faster. To do so, we first determine admissible decay ranges directly from the data. For each contract, all inter-event gaps are pooled and from the $5^{\text{th}}$ and $95^{\text{th}}$ percentiles $(q_{0.05},q_{0.95})$ we define half-life bounds:
\[
h_{\min}=\max(10^{-4},\,0.5\,q_{0.05}), \qquad
h_{\max}=\max(1,\,5\,q_{0.95}),
\]
which translate to exponential decay bounds:
$\beta_{\min}=\log(2)/h_{\max}$ and $\beta_{\max}=\log(2)/h_{\min}$.

Baselines are seeded from empirical rates. For each event type $e$, the total observation time $T$ and event counts are computed, and the initial baseline is set as:
\[
v^{(0)}_{x,e}
=\max\!\Big(10^{-6},\ 
\frac{\#\{\text{type-$e$ events}\}}{T}\Big),
\]
scaled by state: $v^{(0)}_{0,e} = 0.5\,\hat{\nu}_e$ for the Low state and $v^{(0)}_{1,e} = 1.5\,\hat{\nu}_e$ for the Normal state, where $\hat{\nu}_e$ is the empirical rate of event type $e$.

Decay rates $\beta_{x,\,e'\rightarrow e}$ are initialized on a geometric ladder spanning the admissible range $[\beta_{\min},\,\beta_{\max}]$. For the single-exponential specification ($M=1$), this reduces to the geometric midpoint $\beta^{(0)}_{x,e'\rightarrow e}=\sqrt{\beta_{\min}\beta_{\max}}$. Excitation amplitudes are initialized proportionally as:
\[
\alpha^{(0)}_{x,\,e'\rightarrow e,\,m} = \frac{0.1}{M}\,\beta^{(0)}_{x,\,e'\rightarrow e,\,m},
\]
which targets an initial branching ratio of $0.1/M$ per kernel component, ensuring that the starting spectral radius remains well below the instability threshold.

Minimization of the penalized likelihood \eqref{eq:objective} is carried out with the L-BFGS-B quasi-Newton algorithm under box constraints:
$v_{x,e} \in [10^{-4},\,\max(100\,v^{(0)}_{x,e},\,10)]$,
$\alpha_{x,\,e'\rightarrow e}\in [10^{-4},\,500]$,
and $\beta_{x,\,e'\rightarrow e}\in[\beta_{\min},\,\beta_{\max}]$,
where $\beta_{\min} = 0.01$ and $\beta_{\max} = 2500$ are the global bounds applied after the data-driven range estimation. To mitigate sensitivity to the starting point, the optimization is repeated over $n=3$ independent runs: the first uses the data-driven initialization described above, while subsequent runs apply multiplicative jitter drawn uniformly from $[0.5, 1.5]$ to the initial parameters within the bounds. The solution with the lowest objective value is retained.

\end{document}